\documentclass[fleqn,usenatbib]{mnras}
\usepackage{newtxtext,newtxmath}
\usepackage[T1]{fontenc}
\usepackage{ae,aecompl}

\usepackage{graphicx}	% Including figure files
\usepackage{amsmath}	% Advanced maths commands
\usepackage{booktabs} % For prettier tables
\usepackage{subfig}
\usepackage[shortlabels]{enumitem}
\setlist[itemize]{leftmargin=5.5mm}

\newcommand{\dd}[1]{\mathrm{d}#1}
\newcommand{\xv}{\mathbf{x}}

\newcommand{\kv}{\mathbf{k}}

\newcommand{\pv}{\mathbf{p}}
\newcommand{\qv}{\mathbf{q}}

\newcommand{\Pl}{P_{\mathrm{L}}}

\newcommand{\kF}{k_{\rm F}}

\newcommand{\chim}{\chi^2_\mathrm{m}}
\newcommand{\chis}{\chi^2_\mathrm{s}}
\newcommand{\kmax}{k_\mathrm{max}}
\newcommand{\kfit}{k_\mathrm{fit}}
\newcommand{\Fn}{F_n}
\newcommand{\deltainit}{\delta_\mathrm{L}}

\newcommand{\be}{\begin{equation}}
\newcommand{\ee}{\end{equation}}
\newcommand{\bea}{\begin{eqnarray}}
\newcommand{\eea}{\end{eqnarray}}
\newcommand{\bdm}{\begin{displaymath}}
\newcommand{\edm}{\end{displaymath}}

\usepackage[normalem]{ulem}
\usepackage{color}
\definecolor{darkgreen}{RGB}{0,128,0}
\definecolor{darkorange}{RGB}{213,113,0}

\newcommand{\Mpc}{\, h^{-1} \, {\rm Mpc}}

\newcommand{\cGpc}{\, h^{-3} \, {\rm Gpc}^3}
\newcommand{\kMpc}{\, h \, {\rm Mpc}^{-1}}

\newcommand{\eftir}{IR-resummed EFT}
\newcommand{\minerva}{\textsc{Minerva} }
\newcommand{\eos}{\textsc{Eos} }

\title[The matter bispectrum]{The reach of next-to-leading-order perturbation theory for the matter bispectrum}
\author[D. Alkhanishvili et al.]
  {Davit Alkhanishvili,$^1$ %\thanks{Affiliated to ICRA.}
  Cristiano Porciani,$^1$
  Emiliano Sefusatti,$^{2,3,4}$ Matteo Biagetti,$^{2,3,4,5}$
  \newauthor
  Andrei Lazanu,$^{6}$
  Andrea Oddo,$^{3,5}$ and
  Victoria Yankelevich$^{7}$\\
$^1$Argelander Institut für Astronomie der Universität Bonn, Auf dem Hügel 71, 53121 Bonn, Germany\\
$^2$ Istituto Nazionale di Astrofisica, Osservatorio Astronomico di Trieste, via Tiepolo 11, 34143 Trieste, Italy\\
$^3$ Institute for Fundamental Physics of the Universe, via Beirut 2, 34151 Trieste, Italy\\
$^4$ Istituto Nazionale di Fisica Nucleare, Sezione di Trieste, Via Valerio 2, 34127 Trieste, Italy\\
$^5$SISSA - International School for Advanced Studies, Via Bonomea 265, 34136 Trieste, Italy \\
$^6$ Laboratoire de Physique de l'Ecole normale sup\'erieure, ENS, Universit\'e PSL, CNRS, Sorbonne Universit\'e, Universit\'e de Paris, F-75005 Paris, France \\
$^7$ Astrophysics Research Institute, Liverpool John Moores University, Liverpool, L3 5RF, UK \\
}

\date{Accepted XXX. Received YYY; in original form ZZZ}

\pubyear{2021}

\begin{document}
\label{firstpage}
\pagerange{\pageref{firstpage}--\pageref{lastpage}}
\maketitle

\begin{abstract}
We provide a comparison between the matter bispectrum 
derived with different flavours of perturbation theory at next-to-leading order and measurements from an unprecedentedly large
suite of $N$-body simulations. We use
the $\chi^2$ goodness-of-fit test to
determine the range of accuracy of the models as a function of the volume covered by subsets of the simulations.
We find that models based on the effective-field-theory (EFT) approach have the largest reach, standard perturbation theory has the shortest, and `classical' resummed schemes lie in between. 
The gain from EFT, however, is less
than in previous studies.
We show that the estimated range of accuracy of the EFT predictions is heavily influenced by the procedure adopted to fit the amplitude of the counterterms. For the volumes 
probed by galaxy redshift surveys, our results indicate that it is advantageous to set three counterterms of the EFT bispectrum to zero and measure the fourth from the power spectrum. We also find that large fluctuations in the estimated reach occur between
different realisations.
We conclude that
it is difficult to unequivocally define a range of accuracy for the models containing free parameters. Finally, we approximately account for systematic effects introduced by the $N$-body technique  
either in terms of a scale- and shape-dependent bias or by boosting
the statistical error bars of the measurements (as routinely done in the literature). We find that the latter approach artificially inflates the reach of EFT models due to the presence of tunable parameters. 
           
\end{abstract}


%%%%%%%%%%%%%%%%%%%%%%%%%%%%%%%%%%%%%%%%%%%%%%%%%%

%%%%%%%%%%%%%%%%% BODY OF PAPER %%%%%%%%%%%%%%%%%%

\section{Introduction}

The three-point correlation function or its Fourier counterpart, the bispectrum, are the lowest-order clustering statistics that characterise departures from Gaussianity
in the galaxy distribution. Although
these statistics have a long history dating back to the earliest galaxy redshift surveys
\citep{PeeblesGroth1975, Gaztanaga1994, ScoccimarroEtal2001, VerdeEtal2002,  GaztanagaEtal2005, PanSzapudi2005, KulkarniEtal2007, McBrideEtal2011,GilMarin2015,GilMarin2015b,GilMarin2017,SlepianEtal2017a,SlepianEtal2017,Slepian+2018,Pearson2018,Gualdi+2019}, their importance has always been rather marginal.
On the contrary,
they are expected to play a key role to fully exploit the potential of forthcoming observations such as
those conducted with the Dark-Energy Spectroscopic Instrument
\citep[DESI,][]{DESI2016} and the Euclid satellite \citep{Euclid2011}
 by improving
the estimation of cosmological parameters and breaking degeneracies that emerge from the analysis of the power spectrum \citep[e.g][]{ Yankelevich2019, ChudaykinIvanov2019, HeinrichDore2020, Barreira2020, Gualdi+2020,MoradinezhadEtal2020, EggemeierEtal2021,Hahn+2021, Nishant+2021,SamushiaSlepian2021}.

In order to achieve this goal, it is crucial that accurate theoretical models are available in the mildly non-linear
regime of perturbation growth.
The information we want to retrieve, in fact, is distributed over many
triangular configurations the number of which grows rapidly with the minimum considered length scale 
\citep{SefusattiScoccimarro2005, Sefusatti2006, Chan2017}. Therefore, an essential feature of the models is that they give accurate predictions over the widest possible range of scales.
In this paper, we investigate the accuracy of perturbative models
for the bispectrum of the matter density field against large suites of $N$-body simulations. This way we test
the primary building block of
models for the galaxy bispectrum 
that should also address additional
sources of non-linearities (e.g. galaxy biasing, redshift-space distortions) and discreteness effects.

Standard perturbation theory \citep[SPT, see][for a review]{Bernardeau2002} has long been the workhorse for theoretical
predictions on clustering statistics 
in cosmology. Based on the so-called single-stream approximation (in which velocity dispersion is neglected), 
it expands the fluid equations for a self-gravitating pressureless fluid
in terms of the linear density contrast and velocity potential.
The evolution of the bispectrum 
generated 
from Gaussian initial conditions 
to the lowest non-vanishing order
in SPT
was pioneered by \citet{Fry1984}.
Next-to-leading-order (NLO) 
corrections were then discussed in \citet{1997ApJ...487....1S} and \citet{Scoccimarro1998}.
Over the years, alternative
schemes have been developed to model
the growth of cosmological perturbations
and the calculation of the matter bispectrum has been 
combined with techniques that
resum infinite subsets of perturbative contributions
in both the Eulerian~\citep{2006PhRvD..73f3519C,2006PhRvD..73f3520C,Bernardeau2008,PhysRevD.85.123519,Crocce2012} and the Lagrangian descriptions~\citep{Matsubara2008, Rampf2012b}. 
More recently, 
following a general trend in theoretical physics, an effective field theory (EFT)
that approximately describes gravitational instability on `perturbative' scales 
and averages over small-scale fluctuations provided a new framework
to model the matter bispectrum \citep{Angulo2015,Baldauf2015}. In this case, feedback from the small scales to the large scales is expressed in terms of a number of parameters that are fit to observational data or numerical simulations. This construction appears to be successful in extending the range of accuracy of the models. In this work, we compare the different approaches ranging from SPT to EFT with two large suites of $N$-body simulations.
Since many of the forthcoming observational probes will concentrate on intermediate redshifts, 
we only consider data at redshift $z=1$.

Although several authors already tried to determine the
domain of accuracy (sometimes dubbed the reach or $k$-reach) of different perturbative predictions for the bispectrum \citep[e.g.][]{Angulo2015, Baldauf2015b, Lazanu2016, SteeleBaldauf2020}, our work critically evidences that the results depend on a number of factors that have been rarely
explored in depth. In the first place, they depend on the overall volume covered by the $N$-body simulations which determines the size of the statistical uncertainty affecting the measurements. Besides, when these random errors are small, estimates of the reach
are influenced by systematic shifts
due to imperfections of the $N$-body technique
(which are not easy to model and to account for). Furthermore, in the case of the EFT, on top of the
sheer goodness-of-fit criterion one should also consider the consistency (as a function of the minimum length scale under study) of the best-fitting values for the parameters that determine the amplitude of the EFT corrections.
In addition, the number of free parameters and the range of scales 
used in the fitting procedure might
influence the inferred range of accuracy.
By considering all these effects,
we provide a much more comprehensive
investigation of the reach of perturbative models for the matter bispectrum at NLO than what is already available in the literature. As a byproduct of our study, we also obtain analogous results for the matter power spectrum which we use in order to calibrate the EFT corrections for the bispectrum.

This paper is organized as follows. In section~\ref{sec:theory}, we briefly review the PT models we use while, in section~\ref{sec:nbody_sims}, we introduce the simulation suites and the bispectrum measurements. In section \ref{sec:implementation_of_models}, we describe how the perturbative models are implemented in practice
and how we use
the $\chi^2$ goodness-of-fit test to determine their range of accuracy. Our results are presented and critically discussed in section~\ref{sec:results}. Finally, we summarise our findings in
section~\ref{sec:conclusion}.

\section{Perturbation theory}
\label{sec:theory}

Given the Fourier transform of the
mass-density contrast at redshift $z$, $\delta(\kv,z)$,
the matter power spectrum, $P(k,z)$,
and the bispectrum, $B(\kv_1,\kv_2,\kv_3,z)$,
can be defined in terms of the two- and three-point equal-time correlators as
\begin{align}
\label{eq:powerspectrum_definition}
\langle\delta(\kv_1,z)\,\delta(\kv_2,z)\rangle &= (2\pi)^3 \delta_\mathrm{D}(\kv_{12})\, P(k_1,z) \, ,
\end{align}
and 
\begin{align}
\label{eq:bispectrum_definition}
\langle\delta(\kv_1,z)\,\delta(\kv_2,z)\, \delta(\kv_3,z)\rangle = (2\pi)^3 \delta_\mathrm{D}(\kv_{123})\,  B(\kv_1,\kv_2,\kv_3,z) \, ,
\end{align}
where the angle brackets denote averaging over an ensemble of realisations, $\delta_{\rm D}$ is the Dirac delta function, and $\kv_{i\dots j}\equiv\kv_i+\dots+\kv_j$. 
In this section, we briefly review a number of perturbative methods that have been used to model $P(k,z)$ and
$B(\kv_1,\kv_2,\kv_3,z)$
on quasi-linear scales and that we will
test against numerical simulations.

\subsection{Standard perturbation theory}
\label{sec:spt}
In the standard model of cosmology,
the formation of the large-scale 
structure of the Universe is dominated by a dark-matter component. Although the physical origin of dark matter is still unclear, it is generally assumed that, on macroscopic scales, it can be modelled as a self-gravitating medium
governed by the collisionless Boltzmann (or Vlasov) equation in a cosmological background.
The Vlasov-Poisson system can be written as a hierarchy of coupled evolution
equations for the velocity moments
of the phase-space distribution function (the so-called macroscopic transport equations).

SPT assumes that the dark matter 
can be treated as a pressureless ideal fluid governed by the continuity, Euler, and Poisson equations. These are obtained by setting to zero the second velocity moment of the phase-space distribution function and
thus correspond to considering the so-called `single-stream' regime in which there is a well defined velocity everywhere.

By considering irrotational flows only, the dynamic equations are written
in terms of two scalar fields, namely the matter-density contrast and the divergence of the peculiar velocity.
The solution of the linearised transport
equations is 
$\delta^{(1)} (\kv,z)\equiv D(z)\,\deltainit(\kv)$ where $D(z)$ denotes the linear growth factor and $\deltainit(\kv)$ is the linear
solution at the time in which $D=1$. Following an established practice, we set $D=1$ at the present time, corresponding to $z=0$.

The fastest growing solution for $\delta (\kv,z)$ is written as an expansion in terms of the linear density contrast. In particular, if we consider
the Einstein-de Sitter (EdS) cosmological model, it follows that
\begin{equation}
\label{eq:spt_expansion}
\delta(\kv,z)=\sum_{n=1}^{\infty}[D(z)]^n\,\delta^{(n)}(\kv) \, ,
\end{equation}
with 
\begin{align}
\label{eq:psi_n}
\nonumber
\delta^{(n)}(\kv) = \int \frac{\dd^3\mathbf{k}_1 \cdots \dd^3\mathbf{k}_n}{(2\pi)^{3(n-1)}}\, \delta_\mathrm{D}(\mathbf{k}-&\kv_{1\cdots n})\,\Fn(\mathbf{k}_1,\dots,\mathbf{k}_n) \\ &\times\deltainit(\mathbf{k}_1)\cdots\deltainit(\mathbf{k}_n) \, ,
\end{align}
where the (symmetrised) kernels $\Fn$ describe the gravitational coupling between Fourier modes of the linear solution and can be obtained by recursion relations~\citep{GoroffEtal1986}. 

Once the statistical properties of 
$\deltainit(\mathbf{k})$ have been specified, the expressions above allow us to derive perturbative expansions for the power spectrum and the bispectrum of
the matter density contrast.
These are conveniently computed by
following a diagrammatic approach which
is analogous to the Feynman diagrams
in quantum electrodynamics \cite[e.g.][]{Bernardeau2002}.
We classify as `tree-level' all terms associated with tree diagrams (in the sense of graph theory) and as
`loop corrections' those
associated with diagrams containing
$n$-loops (and that require $n$ three-dimensional integrations).

Under the assumption that $\deltainit(\mathbf{k})$ is a Gaussian
random field, 
the leading-order term for $P(k,z)$ coincides with the linear power spectrum, $P^\text{tree}_\text{SPT}(k,z)=[D(z)]^2\,\Pl(k)$, where $\langle \delta_\mathrm{L}(\kv)\,\delta_\mathrm{L}(\kv^\prime) \rangle = (2\pi)^3 \delta_\mathrm{D}(\kv + \kv^\prime) \,\Pl(k)$,
while, for the bispectrum, we have
\begin{align}
\label{eq:bispectrum_tree_level}
\nonumber
B^\text{tree}_\text{SPT}(\kv_1,\kv_2,\kv_3,z)=2 \,[D(z)]^4\,  F_2(\kv_1,\kv_2)\,\Pl(k_1)\,&\Pl(k_2) 
\\ &
+2 \, \text{perms} \,.
\end{align}
Accounting for the NLO corrections, we obtain
\begin{align}
    P_\text{SPT}(k,z)\simeq P^\text{tree}_\text{SPT}(k,z)+P^\text{1-loop}_\text{SPT}(k,z) \, ,
\end{align}
\begin{align}
    B_\text{SPT}(\kv_1,\kv_2,\kv_3,z)\simeq B^\text{tree}_\text{SPT}(\kv_1,&\kv_2,\kv_3,z)\nonumber\\ &+B^\text{1-loop}_\text{SPT}(\kv_1,\kv_2,\kv_3,z) \, .
\end{align}
Their explicit expressions are given in Appendix \ref{appendix:spt}.

On large scales and at early times, tree-level SPT provides an accurate description of both the matter power spectrum and bispectrum. 
At late times, however, 
one-loop corrections over-predict $P$ on mildly non-linear scales \citep[$k\sim 0.1\kMpc$,][]{2006PhRvD..73f3519C,2009PhRvD..80d3531C,2009PhRvD..80l3503T} and
higher-order terms do not improve
the quality of the predictions \citep[e.g.][]{Blas2014}.
The reason for the breakdown of SPT is well understood: loop integrals extend to scales at which the assumptions of the theory do not apply \citep [e.g. due to the generation of vorticity and velocity dispersion at orbit crossing,][]{PueblasScoccimarro2009} and physics becomes non-perturbative. The failure of SPT on small scales thus
corrupts its predictions for the large scales.

%%%%%%%%%%%%%%%%%%%%%%%%%%%%%%%%%%
\subsection{Renormalised perturbation theory}
\label{sec:mpp_model}

Higher-order SPT corrections in 
the expansions for the matter power spectrum and the bispectrum 
may have larger amplitudes than lower-order ones.
In other words, increasing the order of the expansions does not
necessarily improve their accuracy \citep{2006PhRvD..73f3519C, Blas2014}.
Renormalised perturbation theory \citep[RPT,][]{2006PhRvD..73f3519C,2006PhRvD..73f3520C,Crocce2008,Bernardeau2008,PhysRevD.85.123519,Crocce2012} forms one of the first attempts to overcome the shortcomings of SPT \citep[for an approach based on the renormalisation group see][]{MatarresePietroni2007, Pietroni2008}. In RPT, infinite subsets of SPT diagrams are resummed and organized in terms of multi-point propagators defined as the ensemble average of the infinitesimal variation of the evolved cosmic fields with respect to the linear solutions (see Appendix \ref{appendix:rpt}). A key property is that all the statistical quantities such as the power spectra and the bispectra can be expressed in terms of the multi-point propagators. This is known as the multi-point-propagator expansion or $\Gamma$-expansion.

RPT has two main advantages over SPT. First, all the contributions to the power spectrum are positive and adding higher-order terms
improves the range of accuracy of the theory as
no cancellations occur between successive loop corrections. 
Second, the exponential factor appearing in the high-$k$ limit of the multi-point propagators effectively
damps the contributions to the loop integrals outside the range of validity of the expansion, thus
preventing some of the issues which occur in SPT.

One can construct a matching scheme for any multi-point propagator which smoothly interpolates between the resummed behaviour in the high-$k$ limit and the SPT results at low $k$ 
\citep{PhysRevD.85.123519, Crocce2012, Taruya2012}. 
In this paper, we adopt the form derived
in \citet{Taruya2012} which is known as
regularised PT (\textsc{RegPT}).
An alternative matching scheme (dubbed \textsc{MPTbreeze}) has been proposed by \citet{Crocce2012} and implemented for the bispectrum in \citet{Lazanu2016}.
We have verified that \textsc{RegPT} and \textsc{MPTbreeze} give nearly identical results and, for this reason, there is no point in considering both here.
%%%%%%%%%%%%%%%%%%%%%%%%%%%%
\subsection{Lagrangian perturbation theory}

In the Lagrangian approach to fluid dynamics, the trajectories of the fluid elements are characterized in terms 
of the displacement field $\mathbf{\Psi}(\pv, t)$ which 
links
the Lagrangian position $\pv$ and the Eulerian position $\xv$ (at time $t$) through the relation
$\xv(\pv, t) = \pv + \mathbf{\Psi} (\pv, t )$. Lagrangian perturbation theory (LPT) is derived by using 
$\mathbf{\Psi} (\pv, t )$ as a perturbative variable
\cite[e.g.][]{Zeldovich1970, Moutarde1991, Catelan1995}.
In this framework, 
the Eulerian matter density can be expressed as 
\begin{equation}
\delta (\mathbf{k}) = \int \dd^3 \mathbf{p} \, e^{-i \mathbf{k} \cdot \mathbf{p}} \left[ e^{-i \mathbf{k} \cdot \mathbf{\Psi} (\mathbf{p}) }  -1 \right] \, ,
\end{equation}
(where we do not write the time dependence explicitly to simplify notation)
which allows us to write an expression for the power spectrum 
\begin{equation}  \label{eq:RLPT_P1}
P(\mathbf{k}) = \int \dd^3 \mathbf{\Delta}_{12} \, e^{-i \mathbf{k} \cdot \mathbf{\Delta}_{12}} \left[ \langle e^{-i \mathbf{k} \cdot \left[  \mathbf{\Psi}(\mathbf{p_1}) -  \mathbf{\Psi}(\mathbf{p_2}) \right] }  \rangle - 1 \right] \, ,
\end{equation}
and the bispectrum 
\begin{align} \label{eq:RLPT_B1}
\nonumber
B(\kv_1,\kv_2,\kv_3) &= \int \dd^3 \mathbf{\Delta}_{12} \int \dd^3 \mathbf{\Delta}_{13} \, e^{-i \mathbf{k} \cdot \left(\mathbf{\Delta}_{12} + \mathbf{\Delta}_{13} \right) } \\
 \times &\left[ \langle e^{-i \mathbf{k_2} \cdot \left[ \mathbf{\Psi}(\mathbf{p_1}) -  \mathbf{\Psi}(\mathbf{p_2})   \right]   -i \mathbf{k_3} \cdot \left[ \mathbf{\Psi}(\mathbf{p_1}) -  \mathbf{\Psi}(\mathbf{p_3})   \right] }  \rangle  - 1  \right] \, ,
\end{align}
where $ \mathbf{\Delta}_{ij} \equiv  \mathbf{p_i} -  \mathbf{p_j} $, and the expectation value only depends on the separation $\mathbf{\Delta}_{12}, \mathbf{\Delta}_{13}$ due to homogeneity \citep{Taylor1996,Fisher1996,Matsubara2008, Rampf2012b}.
A perturbative expansion of equations~(\ref{eq:RLPT_P1}) and (\ref{eq:RLPT_B1}) can then be obtained by means of the cumulant expansion theorem 
\begin{equation}
\langle e^{-iX} \rangle = \exp {\left[  \sum_{N=1}^{\infty}  \frac{(-i)^N}{N!} \langle X^N  \rangle_c  \right]} \, ,
\end{equation}
where $\langle X^N  \rangle_c$ represents the $N^\mathrm{th}$ order cumulant of the random variable $X$. Expanding the powers of $X$ with the binomial theorem, two types of terms are obtained: those depending on $\mathbf{\Psi}$ at one point, and those depending on $\mathbf{\Psi}$ at multiple points. It turns out that,
 if both sets of terms are expanded to the same perturbative order, the `classical' LPT results coincide with the SPT expressions for both the power spectrum and the bispectrum \citep{Matsubara2008, Rampf2012b}.

\subsection{Resummed Lagrangian perturbation theory}
On closer inspection, it emerges the classical LPT predictions for $P$ and $B$ can be improved by reorganising the
perturbative expansion.
The key issue is that, for large Lagrangian separations, the terms depending on $\mathbf{\Psi}$ at one point are much larger than those depending on $\mathbf{\Psi}$ at multiple points.
It thus makes sense to keep the first set of terms inside the argument of the exponential and use the cumulant expansion only for the second set \citep{Matsubara2008, Rampf2012b}. 
This approach is generally referred to as resummed Lagrangian perturbation theory (RLPT) as 
it corresponds to a partial resummation of the perturbative expansion\footnote{Note that RLPT is different from the so-called
convolution Lagrangian perturbation theory \citep[e.g.][]{2013MNRAS.429.1674C} which further extends the partial resummation but has not yet been applied to the bispectrum.}.
The resulting formulae for calculating
$P$ and $B$ to NLO are presented in
Appendix \ref{appendix:rlpt}.
In \citet{Lazanu2016}, it has been shown that the RLPT predictions are similar to those of \textsc{MPTbreeze}.

\subsection{Effective Field Theory of Large-Scale Structure}
\label{sec:EFT}

Effective theories have become a widely used tool in modern physics.
In a system characterised by a wide range of scales, 
they isolate a set of degrees of freedom and
describe them with a simplified model
without having to deal with the complex (and often unknown) underlying dynamics.
The impact of the physics one wishes to neglect on the degrees of freedom one desires to study is
computed as a perturbation theory in terms of one or more expansion parameters.

The effective field theory of large-scale structure \citep[EFT,][]{BaumannEtal2012, Carrasco2012,Carrasco2014,Carrasco2014b,Hertzberg2014,Porto2014, SenatoreZaldarriaga2015} attempts to provide an effective description of the long-wavelength modes of the matter density field by integrating out (i.e. averaging over)
the short-wavelength ones. Contrary to the models introduced in the previous sections,
the EFT does not rely on the single-stream
approximation and considers an effective
stress tensor which is expressed
in terms of all operators of the long-wavelength density and velocity fields (and their derivatives) allowed by the symmetries of the problem: the equivalence principle along with the assumption of statistical isotropy and homogeneity. 
The effective stress tensor is Taylor expanded in the long-wavelength fluctuations
giving rise to
an infinite series of unknown
parameters each associated with a perturbative order.
These parameters can be treated as coupling constants in the Wilsonian approach to renormalisation.
We can imagine that the theory contains
a cutoff (i.e. the loop corrections
are integrated up to a maximum wavenumber)
and the couplings of the effective theory can be changed to enforce that the physics at low $k$ is always the same when the cutoff is changed.
Therefore, the parameters of the effective theory fulfil two purposes.
In the EFT expressions for observables, 
they generate `counterterms' which can be used to cancel out the UV sensitivity of the loop integrals in SPT (i.e. their dependence on the cutoff scale). 
This can be done order by order in perturbation theory.
Moreover,
the remaining cutoff-independent part of the counterterms should actually quantify the impact of the non-perturbative physics on the long-wavelength modes by introducing new `effective' interactions among long-wavelength modes.
The amplitude of this part, however, cannot be derived from the EFT
(which is blind to small-scale physics) and must be fixed empirically by comparison with numerical simulations or marginalised over in the analysis of actual observational data \citep[see, e.g.][]{IvanovEtal2020, DAmicoEtal2020}.

EFT assumes the existence of a scale, generally indicated in terms of the wavenumber $k_{\mathrm{NL}}$, around which physics becomes non-perturbative and the effective description becomes meaningless. Several lines of reasoning
suggest that the derivative expansion
of the long-wavelength fields can be
organised
so that the expansion parameter of the perturbation theory
is $k/k_{\mathrm{NL}}$, meaning that more and more terms should be considered
to get accurate expressions for the
correlators of the matter field
as $k$ approaches $k_{\mathrm{NL}}$.

The fact that perturbations of all wavelengths (barring virialised structures) evolve
on similar time scales
constitutes a complication of the theory. It follows from this that the EFT is non-local in time, i.e. the long-wavelength perturbations depend on the entire past history of the short-wavelength modes. This is difficult to treat and, in practical applications, the local-in-time approximation is almost invariably invoked. We adopt the same strategy in our study.
In particular, we focus on the specific parameterisation of the counterterms appearing in the one-loop expressions for the matter power spectrum and bispectrum presented in \citet{Angulo2015}. Considering the linear Taylor approximation of the effective stress tensor in the long-wavelength perturbations
gives the EFT power spectrum to NLO
\citep{Carrasco2014b}
\begin{equation}
\label{eq:eftP0}
P_\text{EFT} (k, z) = P_\text{SPT} (k, z) + P_{c_0} (k, z) \, ,
\end{equation}
where the tree-level counterterm is given by
\begin{equation} \label{eq:eftP}
P_{c_0} (k, z) = -2\,c_0(z)\,[D(z)]^2\,k^2\,\Pl\left(k\right) \, ,
\end{equation}
and $c_0$ is undetermined by the theory.
In terms of 
the effective speed of sound for the perturbations, $c_{{\mathrm s}(1)}(z)$,
we have
$\bar{c}_0\equiv (2\pi)\,[D(z)]^{\zeta}\, [c_{\mathrm{ s}(1)}(z)]^2/k_\mathrm{NL}^2$.
Note that our $c_0$ relates
to the parameter $\bar{c}_1$
introduced by \citet{Angulo2015} as
$c_0\equiv \bar{c}_1\,[D(z)]^{\zeta}$
where $[D(z)]^{n+\zeta}$ is the assumed growth factor of the EFT corrections to the SPT density fluctuations of order $n$.

Similarly, for the bispectrum to NLO, EFT gives four counterterms \citep{Angulo2015, Baldauf2015b}
\begin{align}
B_\text{EFT}= 
B_\text{SPT}+B_{c_0}+B_{c_1}
+B_{c_2}+B_{c_3}\;.
\end{align}
(where the dependence
on $\kv_1,\kv_2,\kv_3$ and $z$ is left implicit to simplify notation),
one of which is also proportional to $c_0$
\begin{multline}  \label{eq:eftB}
B_{c_0}=
c_0(z)\,[D(z)]^4\,\left[
2\,\Pl\left(k_1\right)\Pl\left(k_2\right) \tilde{F}_2^{\left(s\right)}\left(\textbf{k}_1,\textbf{k}_2\right) +\text{2 perms.}\right. \\
\left.-2\,k_1^2\Pl\left(k_1\right)\Pl\left(k_2\right) F_2\left(\textbf{k}_1,\textbf{k}_2\right) + \text{5 perms.}\right] \,,
\end{multline}
with
\begin{align}
 \tilde{F}_2^{(s)}(\textbf{k}_1,\textbf{k}_2) &=
	-\frac{1}{(1+\zeta)(7+2\zeta)} \left[ \left( 5 + \frac{113\zeta}{14} + \frac{17\zeta^2}{7} \right) (k_1^2+k_2^2) \right. \nonumber \\
	& + \left( 7 + \frac{148\zeta}{7} + \frac{48\zeta^2}{7} \right) \textbf{k}_1\cdot\textbf{k}_2 + \left( 2+\frac{59\zeta}{7}+\frac{18\zeta^2}{7} \right) \nonumber \\
&  \times\left( \frac{1}{k_1^2}+\frac{1}{k_2^2} \right) 
	(\textbf{k}_1\cdot\textbf{k}_2)^2 + \left( \frac{7}{2}+\frac{9\zeta}{2}+\zeta^2 \right) \left( \frac{k_1^2}{k_2^2} + \frac{k_2^2}{k_1^2} \right) 
\nonumber \\
&\left.\times\, \textbf{k}_1\cdot\textbf{k}_2 
 + \left( \frac{20\zeta}{7} + \frac{8\zeta^2}{7} \right) \frac{(\textbf{k}_1\cdot\textbf{k}_2)^3}{k_1^2 k_2^2} \right] \, .
\end{align}
Following \citet{Angulo2015}, 
we assume $\zeta=3.1$ as suggested by  some theoretical considerations and fits to simulations \citep{Foreman2016}. We note that \citet{Baldauf2015} find no appreciable difference between using
$\zeta=2$ or $3.1$.
Quadratic contributions from the long-wavelength perturbations to the effective-stress-tensor expansion lead to four additional counterterms, only three of which are independent. They have the following forms:
\begin{align}
\label{eq:bc1}
B_{c_1}&= -2\,c_1(z)\,[D(z)]^4\, k_1^2 \Pl(k_2)  \Pl(k_3) + \text{2 perms.}\, , \\ %\nonumber
B_{c_2}&= -2\,c_2(z)\,[D(z)]^4 \, k_1^2 \frac{(\kv_2\cdot\kv_3)^2}{k_2^2 k_3^2} 
	 \Pl(k_2)  \Pl(k_3) \nonumber \\
	 &\ \ \ \ \ \ \ \ \ \ \ \ \ \ \ \ \ \ \ \ + \text{2 perms.}\, , \\ %\nonumber
B_{c_3}&= -c_3(z)\,[D(z)]^4 \, (\kv_2\cdot\kv_3) 
 \left[ \frac{\kv_1\cdot\kv_2}{k_2^2} 
	+ \frac{\kv_1\cdot\kv_3}{k_3^2} \right] 
    \Pl(k_2)  \Pl(k_3)  \nonumber \\
\label{eq:bc3}
    &\ \ \ \ \ \ \ \ \ \ \ \ \ \ \ \ \ \ \ \ + \text{2 perms.} \, , %\\ \nonumber
\end{align}
where the effective coupling constants $c_1$, $c_2$ and $c_3$ are unknown 
(similar to $c_0$, we absorb the 
$[D(z)]^\zeta$ scaling in their definition). 

\begin{table*}
\begin{center}
\begin{tabular}{ccccccccccccc}\hline
    Name & $n_s$ & $h$ & $ \Omega_b$ & $\Omega_m$ & $\sigma_8$ & \# &$N_\mathrm{p}^{1/3}$&$L_{\rm box}$ & $V_\mathrm{tot}$ &$m_\mathrm{p}$ & IC & $z_{\rm initial}$ \\
    & & & & & & sims & &$[\Mpc]$ & $[\cGpc]$ &$[10^{10} \,h^{-1}M_\odot]$ & &  \\
    \hline\hline
    \minerva & $0.9632$&$0.695$&$0.044$&$0.285$&$0.828$&$300$ & $1000$ & $1500$ & $1012$ & $26.7$ & 2LPT & $63$\\
    \hline
    \textsc{Eos} & $0.967$&$0.7$&$0.045$&$0.3$&$0.85$&$10$ & $1536$ & $2000$ & $80$ & $18.3$ & 2LPT & $99$\\
    \hline
\end{tabular}
\caption{Cosmological and structural parameters for the \minerva and \textsc{Eos} simulations.}
\label{tab:sims}
\end{center}
\end{table*}

Although some authors claim that EFT provides a manifestly convergent
perturbative scheme for $k<k_\mathrm{NL}$ \citep[e.g.][]{Carrasco2014},
there are indications that, like SPT, it forms an
%that the perturbative series is an
asymptotic expansion 
in which adding higher and higher-loop corrections, at a certain point, deteriorates
the agreement with numerical simulations
\citep[e.g.][]{Pajer-vanderWoude-2018, Konstandin+2020}.
The break down of the theory should not be caused
by the influence of short-distance physics but rather to large contributions coming from mildly non-linear scales.

At the end of the day, EFT can also be simply seen as an improved version of SPT in which counterterms are added to regularise the UV-sensitive contributions.

\subsection{IR resummation}

Large-scale flows broaden and damp the baryon-acoustic-oscillation (BAO) feature imprinted in $P_\mathrm{L}$ at early epochs.
These effects are poorly captured by
Eulerian perturbation theories and are more easily understood in the Lagrangian framework \citep{MeiksinWhitePeacock1999, Crocce2008, 2009PhRvD..80l3503T}.
It turns out that it is possible to
account for them
by resumming the perturbative predictions to all orders, a procedure known as `IR resummation' \citep[e.g.][]{SenatoreZaldarriaga2015}.
In the framework of EFT, this is often
implemented following the strategy
delineated by \cite{Baldauf-ir-res2015} and further developed in \citet[][the method we use]{Blas2016a} and \citet{Ivanov2018}. 
In order to decompose the linear power spectrum in smooth and oscillating parts, we use one-dimensional Gaussian smoothing as described in \citet[][appendix A]{VlahEtal2016} and \cite{Osato2019}.

\subsection{Time evolution}
\label{sec:timeevo}
In all the results described above, time evolution is entirely captured by the function $D(z)$. This directly follows from equation~(\ref{eq:spt_expansion})
and its analogue for the EFT corrections\footnote{I.e. $\delta_\mathrm{EFT}(\kv,z)=\sum_{n=1}^{\infty}[D(z)]^{n+\zeta}\,\delta_\mathrm{EFT}^{(n)}(\kv)$ where $\delta=\delta_\mathrm{SPT}+\delta_\mathrm{EFT}$.}
which hold true in the EdS universe only. In general, the second-order SPT
solution has the form $D_\mathrm{2A}(z)\,A(\kv)+
D_\mathrm{2B}(z)\,B(\kv)$ 
where $D_\mathrm{2A}(z)$ and $D_\mathrm{2B}(z)$ slightly differ from
$[D(z)]^2$ \citep[for their explicit expressions see e.g. Appendix A in][]{Takahashi2008}.
Similarly,
the third-order solution contains six different growth factors that
deviate a little from $[D(z)]^3$.
Previous studies have shown that assuming the $[D(z)]^n$ scaling provides rather accurate approximations to
the matter power spectrum and bispectrum
in the $\Lambda$CDM model ~\citep[e.g.][]{Scoccimarro1998,Bernardeau2002}.
For $P(k)$, the leading-order contribution is unaffected since it only depends on the linear density fluctuations.
Moreover, in the relevant range of wavenumbers,
deviations from the exact solution for the one-loop corrections are well below the per-cent level at $z=1$ \citep{Takahashi2008}. 
For these reasons, we can safely set $D_\mathrm{2A}(z)=D_\mathrm{2B}(z)=D(z)$ in our analysis of the power spectrum.
On the contrary, we use the exact $F_2$ kernel 
\begin{align}
F_\mathrm{2,\Lambda CDM}(\kv_1,\kv_2)=
\nonumber
\frac{5}{7}&\frac{D_{2A}(z)}{D(z)^2}\frac{(\kv_1+\kv_2)\cdot\kv_1}{k_1^2}\\
&+\frac{2}{7}\frac{D_{2B}(z)}{D(z)^2}\frac{(\kv_1+\kv_2)^2\,\kv_1\cdot\kv_2}{2 k_1^2 k_2^2}\,,
\end{align}
to compute the tree-level bispectrum in SPT and EFT (but not for the loop corrections).
This is necessary because adopting the EdS approximation would generate systematic shifts at the per-cent level \citep{SteeleBaldauf2020} which are comparable with the statistical errors of the measurements extracted from our very large suites of simulations (see section~\ref{sec:nbody_sims}).
We revisit this issue in section~\ref{sec:bisp}.

Since we only consider the matter density field at $z=1$, from now on, we drop the dependence on $z$ of all functions.

\section{$N$-body simulations}
\label{sec:nbody_sims}
In this section, we introduce the $N$-body simulations and the estimators we use to test the theoretical models introduced above.
\subsection{Simulation suites}
We use two sets of $N$-body simulations, named \minerva and \textsc{Eos},
run using the \textsc{Gadget-2} code \citep{Springel2005}.
Our main investigation is based on the
\minerva set \citep[first presented in][]{Grieb2016} which
consists of 300 simulations each following the evolution of $1000^3$ dark-matter particles in a periodic cubic box with a side length of $1500\Mpc$. 
In order to perform some additional
tests in section~\ref{sec:results},
we complement the \minerva suite with a subset\footnote{
Information on the \textsc{Eos} suite is available in  \citet{Biagetti:2016ywx} and at \texttt{https://mbiagetti.gitlab.io/cosmos/nbody/eos/}.} 
of the \textsc{Eos} suite composed of $10$ realizations
each containing $1536^3$ particles in a periodic cubic box with a side length of $2000\,\Mpc$. 

The simulations follow the formation of the large-scale structure in flat $\Lambda$CDM cosmological models with parameters given in Table \ref{tab:sims}.
The linear transfer  functions are obtained from the Boltzmann codes \textsc{camb}~\citep{Lewis:1999bs, Howlett:2002} and \textsc{CLASS}~\citep{Blas:2011rf} for the \minerva and \eos simulations respectively. In all cases, the initial particle displacements are computed using the publicly available code \textsc{2LPTic} \citep{CroccePueblasScoccimarro2006} starting from Gaussian initial conditions.

%%%%%%%%%%%%%%%%%%%%%%%%%%%%%%%%%%%%%%%%%%%%%%%%%%
%%%%%%%%%%%%%%%%%%%%%%%%%%%%%%%%%%%%%%%%%%%%%%%%%%
\subsection{Power spectrum and bispectrum estimators}
\label{ssec:discreteness}

We use the \textsc{PowerI4} code \citep{Sefusatti+2016} to 
estimate the matter density
in a regular Cartesian grid containing $512^3$ cells
from the particle positions. With the FFT algorithm, we obtain the
Fourier-space overdensity $\delta_\mathbf{q}$ sampled at the wavevectors $\mathbf{q}$ with Cartesian components that are integer multiples of the fundamental frequency $\kF=2\pi/L_\mathrm{box}$.
Our power-spectrum estimator is
\be
\label{eq:power_spectrum_estimator}
\hat{P}(k) = \frac{1}{L_\mathrm{box}^3 N_P}\sum_{\qv \in k} \, |
\delta_\mathbf{q}|^2\,,
\ee
where $N_P$ is the number of $\mathbf{q}$ vectors lying in a bin centred at wavenumber $k$ and of width $\Delta k$. The notation $\qv \in k$ means that $k -\Delta k/2 \le q < k +\Delta k/2 $.
Similarly, for the bispectrum, we use
\be
\label{eq:bispectrum_estimator}
\hat{B}(k_1, k_2, k_3) = \frac{1}{L_\mathrm{box}^3 N_B}\sum_{\qv_1\in k_1} \sum_{\qv_2\in k_2} \sum_{\qv_3\in k_3} \, \delta_{\qv_1} \delta_{\qv_2} \delta_{\qv_3}\,,
\ee
where  $\qv_1$, $\qv_2$ and $\qv_3$ satisfy the triangle condition $\qv_{123}=0$ and $N_B$ denotes the number of triangles contributing to a given `triangle bin' defined by the sides $k_1 \geq k_2 \geq k_3$
\citep[which do not necessarily form a closed triangle,][]{Oddo2019}.
We consider different binning schemes
characterised by the bin width $s=\Delta k/\kF$ and the central wavenumber of the first bin $c$ (also expressed in units of $\kF$) so that the centres of all bins are given by
\begin{equation}
k_i=[c+(i-1)\,s]\, \kF \,,\qquad i=1, 2,\dots, N_k\,.
\label{eq:binning}
\end{equation}
The parameters we use
for the different power-spectrum and bispectrum measurements and the maximum wavenumber they reach are summarised in Table~\ref{tab:binning}. It is worth stressing that  we subtract from $\hat{P}$ and $\hat{B}$ the systematic contributions due to Poissonian shot noise which are anyway smaller than the statistical uncertainties.
\begin{table}
	\centering
	\bgroup
	\def\arraystretch{1}
	\begin{tabular}{l|c|c|c|c}
		\hline
		$s$ & $c$ & $N_k$ & $N_\mathrm{t}$ & $k_\text{max}[\kMpc]$ \\
		\hline
		
		1 & 2.0 & 48 & 11757 & 0.20 \\
		2 & 2.5 & 28 & 2513 & 0.24 \\
		3 & 3.0 & 28 & 2513 & 0.36 \\
		\hline
	\end{tabular}
	\egroup
	\caption{Main characteristics of our binning schemes --see equation~(\ref{eq:binning}). The total number of measurements for $\hat{P}$ and $\hat{B}$ are
	indicated with $N_k$ and $N_\mathrm{t}$, respectively, while
	$k_\mathrm{max}$ gives the maximum
	wavenumber reached.}
	\label{tab:binning}
\end{table}

\begin{figure}
 \centering
 \includegraphics[width=\columnwidth]{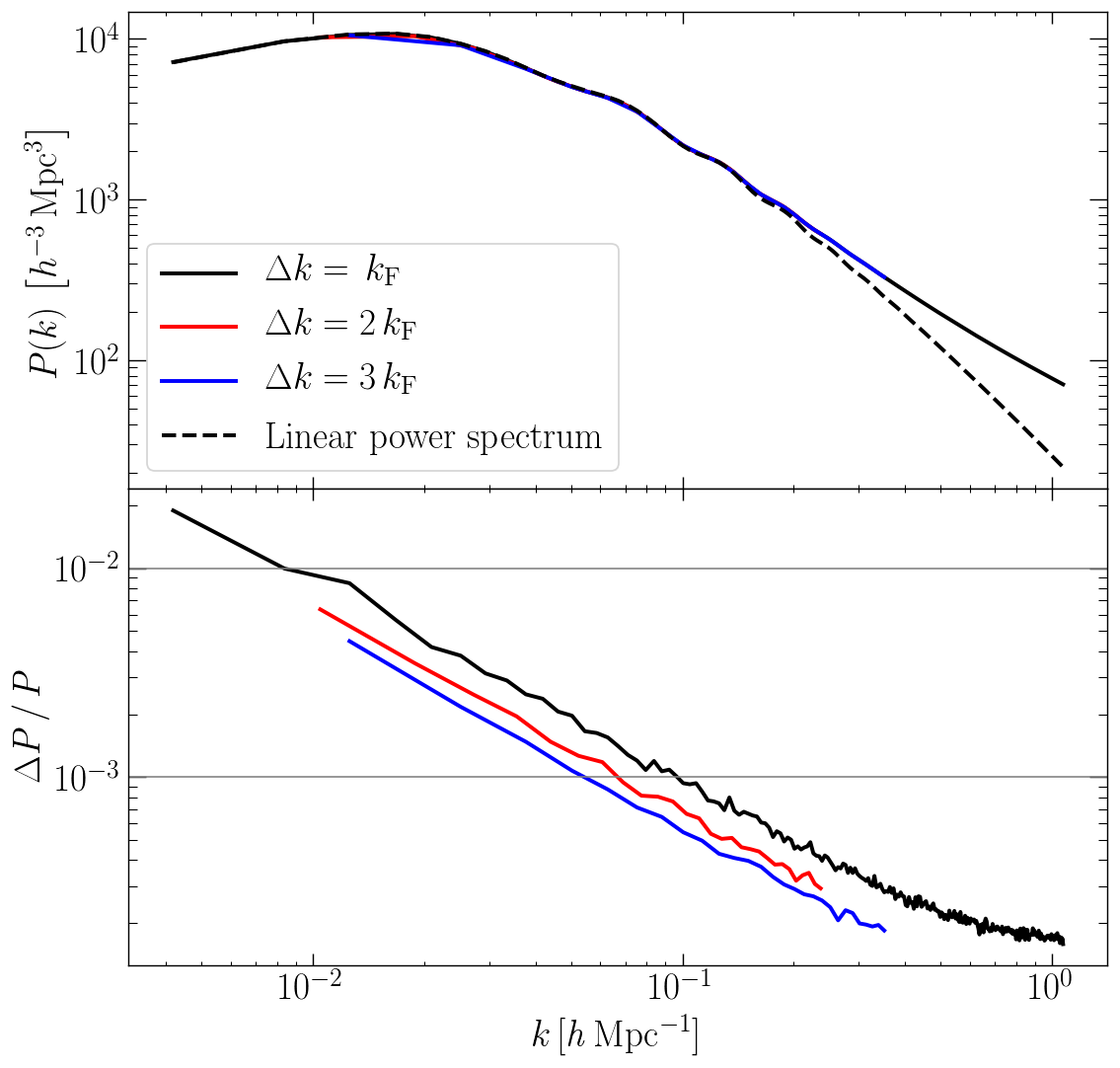}
 \caption{The mean power spectrum extracted from the \minerva simulations (top) and the corresponding statistical uncertainty (bottom).
 }
 \label{fig:ps_numerical}
\end{figure}
\begin{figure*}
 \centering
 \includegraphics[width=.925\textwidth]{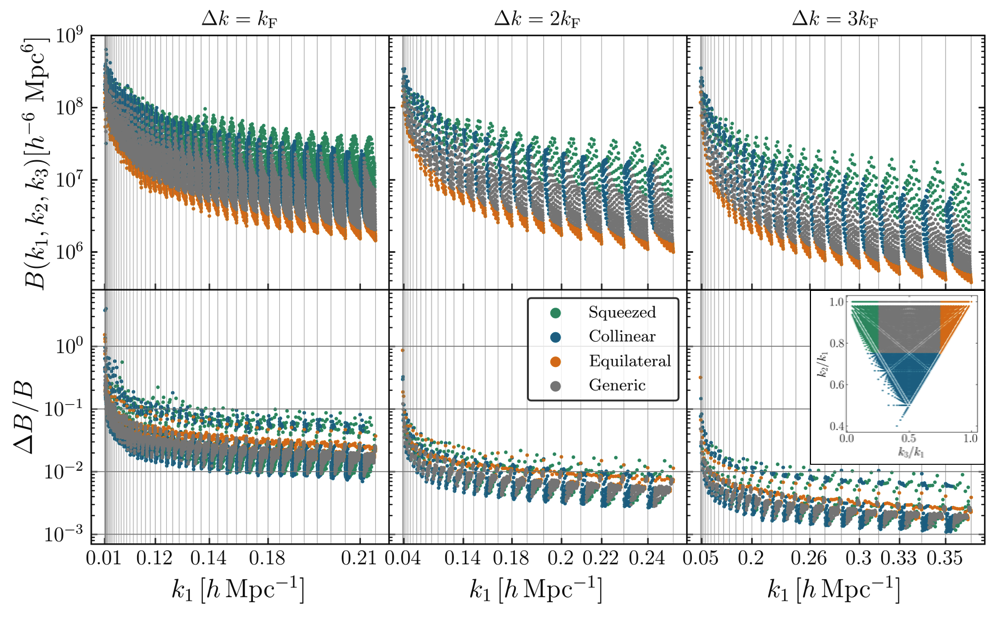}
 \caption{The mean bispectrum
 extracted from the \minerva simulations (top) and its statistical uncertainty (bottom). 
 Results are plotted by ordering the
 triangular configurations as in \citet{Oddo2019}.
 In between two consecutive vertical lines, all points correspond to triangle bins with the same longest side $k_1$, whereas $k_2$ and $k_3$ take all allowed values.
 The color of the symbols indicates different triangular shapes
 as illustrated in the bottom middle
 and right panels.}
 \label{fig:bs_numerical}
\end{figure*}

Fig.~\ref{fig:ps_numerical} shows 
the average $\hat{P}$ obtained from the
\minerva simulations for the three bin sizes (top panel) and the relative standard error of the mean (bottom panel). Note that, due to the large
number of realisations we consider,
we achieve better than one-per-cent (one-per-mille) precision for $k>0.01 \kMpc$ 
 ($k>0.1 \kMpc$).
Similarly, Fig.~\ref{fig:bs_numerical} shows the mean $\hat{B}$ (top panels) and its standard deviation (bottom panels).
In this case, the relative errors
range between 10 per cent and one per mille depending on the triangular
configuration and the bin size.
Dealing with such
unusually small random errors 
(which cannot be obtained from
current observations of galaxy clustering) calls for a consistent treatment of the systematic errors introduced by the $N$-body method (see section~\ref{sec:systematic_errors}).

%%%%%%%%%%%%%%%%%%%%%%%%%%%%%%%%%%%%%%%%%%%%%%%%%%
%%%%%%%%%%%%%%%%%%%%%%%%%%%%%%%%%%%%%%%%%%%%%%%%%%

 %%%%%%%%% Implementation of the models %%%%%%%%%%%%%%%%

 \section{Matching the models to simulations}
 \label{sec:implementation_of_models}

In this section, we explain how we 
compare the perturbative models to 
the measurements extracted from the
$N$-body simulations. 

\subsection{Binning of theoretical predictions}
\label{sec:model_binning}

In order to compare the theoretical predictions with the measurements, we need to account for the finite bin sizes assumed by the power-spectrum and bispectrum estimators and, possibly, for the discreteness characterising the Fourier-space density grid.
The most precise approach to the problem\footnote{With the exception of
point-by-point comparisons on individual realisations \citep{Roth-Porciani-2011,Taruya2012, Taruya+2018,SteeleBaldauf2020}.}  consists of averaging the theoretical predictions over the same set of configurations as it is done for the estimators (\ref{eq:power_spectrum_estimator}) and (\ref{eq:bispectrum_estimator}).
Taking these averages, however, is  computationally demanding, at least for the bispectrum.
A considerable speedup (at the expense of accuracy) can be achieved by computing the model predictions for one characteristic configuration per triangle bin.
For instance, \citet{SefusattiCrocceDesjacques2010} considered the average value of the triplet $(k_1, k_2, k_3)$ in a bin
that from now on we refer to as the
`effective' triangle of a bin.
In what follows, we always use the full average
of the theoretical predictions for the power spectrum and the tree-level bispectrum.
On the other hand,
due to the computational demand,
we average the loop corrections 
for $B$ only for triangle bins with
$k_3\lesssim 0.14\,\kMpc$. In all the other cases,
we evaluate the corrections using one effective triangle per bin (after checking that this approximation is accurate enough on the larger scales for which we have the average).

%%%%%%%%%%%%%%%%%%%%%%%%%%%%%%%%%%%%%%%%%%%%%%%%%%
%%%%%%%%%%%%%%%%%%%%%%%%%%%%%%%%%%%%%%%%%%%%%%%%%%

\subsection{Goodness of fit}
\label{ssec:validation_of_models}
In order to quantify the goodness-of-fit of the
different models, we assume Gaussian errors and rely on the $\chi^2$ test. 
Schematically, given the mean measurements from the simulations, $\langle D_i\rangle$, and the corresponding model predictions, $M_i$, we compute the statistic
\begin{equation}
\label{eq:chi_2}
\chim=\frac{\chi^2_\mathrm{tot}}{\nu}= \frac{1}{\nu}\,\sum_{i,j} (M_i-\langle D_i\rangle)\,C^{-1}_{ij}\,(M_j-\langle D_j\rangle)\, ,
\end{equation} 
where $\nu$ indicates the number of degrees of freedom (i.e. the number 
of data points $N$ minus the number of adjusted parameters),
the indices $i$ and $j$ run over all possible configurations, and $C_{ij}$ denotes the elements of the covariance matrix for the adopted estimators (or some approximation thereof).

Since we only consider relatively large scales,
we use the so-called Gaussian
contribution to the covariance matrix
for the power-spectrum estimates,
\citep{FKP1994,MeiksinWhite1999}
\be
\label{eq:Cpp} 
C_{ij}= \frac{2\,P_i^2}{N_P}\,\delta_{ij}\;,
\ee
with $P_i$ the expected power spectrum
in the $i^\mathrm{th}$ bin and
$\delta_{ij}$ the Kronecker symbol.
In order to prevent that the covariance
is informed about the noise in our realisations, we use a smooth function to compute $P_i$ in the expression above. This is obtained by fitting
the outcome of the \minerva simulations with the expression 
\citep{Cole2005}
\be
\label{eq:P_nl}
P_\mathrm{NL}(k) = \Pl(k)\left( \frac{1+Q k^2}{1+A k}\right )\,,
\ee
where $Q$ and $A$ are free parameters. 
We find that setting
$Q\approx 4\,h^{-2}\,\mathrm{Mpc}^2$ and $A\approx 0.37\,\Mpc$ provides a fit
that agrees with the measurements
to better than one per cent at all the
scales considered in this work.

For the bispectrum,
we find that, even at large scales, the Gaussian approximation underestimates the sample variance from numerical simulations in a shape-dependent manner, reaching a difference of order 50 per cent for some squeezed-triangle configurations \citep[see also][]{Chan2017, Colavincenzo+2019,Gualdi+2020}. 
For this reason, we use the approximate expression
\begin{align}
\label{eq:C_BB}
    C_{ij}&= [(PPP)_i + 2\, (BB)_i]\,\delta_{ij}\;,
\end{align}
where 
\begin{equation}
\label{eq:C_BBGauss}
    (PPP)_i\simeq\frac{6\, L_\mathrm{box}^3}{N_B}\,\overline{ P_\mathrm{NL}(k_1) P_\mathrm{NL}(k_2) P_\mathrm{NL}(k_3)}
\end{equation}
 denotes the Gaussian part and the overline indicates the average
over all the configurations contributing to the $i^\mathrm{th}$ triangle bin while 
\begin{equation}
    (BB)_i\simeq (B_\mathrm{NL}^{\mathrm{eff}})^2\left[\frac{1}{N_P(k_1)}+\frac{1}{N_P(k_2)}+\frac{1}{N_P(k_3)}\right]\,,
\end{equation}
where $B_\mathrm{NL}^{\mathrm{eff}}$ denotes the tree-level bispectrum in SPT evaluated at the effective wavenumbers using $P_\mathrm{NL}$ instead of $P_\mathrm{L}$.
The $(BB)_i$ term approximates the actual non-Gaussian contribution due to
configurations that share one $k$-bin \citep[see, e.g.][]{Sefusatti2006}. Note that the factor of two in equation~(\ref{eq:C_BB}) is meant to
approximately compensate for the additional contribution to the covariance that scales as the product of the power spectrum and the trispectrum.

In order to assess the accuracy of these approximations to the covariance matrices for $\hat{P}$ and $\hat{B}$, we
use the statistic
\begin{equation}
 \chis=\frac{1}{N}\,\mathrm{Tr}(\mathbfss{S}\mathbfss{C}^{-1}) \, ,
\end{equation}
where $\mathbfss{S}$ denotes
the sample covariance matrix
of the measurements from the \minerva simulations and $\mathbfss{C}$ is our
model covariance.
It is possible to show that,
for Gaussian errors with covariance matrix $\mathbfss{C}$, the statistic
$N\chis$ follows a chi-square 
distribution with $N$ degrees of freedom \citep{Porciani-21}. Therefore, our approximations for the covariance matrix should be considered inaccurate if
$\chis$ strongly departs from unity.
In this case, any conclusion on the accuracy of the models based on $\chim$ should be disregarded. 
Note that our approximations for the covariance matrices are diagonal, implying that 
\begin{equation}
\chis=\frac{1}{N}\,\sum_{i=1}^N \frac{S_{ii}}{C_{ii}}\,.
\end{equation}
i.e. $\chis$ gives the average ratio between
the measured and assumed variances of the different data points.

Before moving on, it is important to note that we do not account for the so-called `theoretical errors' -- i.e. uncertainties on the perturbative predictions reflecting the estimated size of the higher-order terms that are neglected -- as advocated by some authors \citep[e.g.][]{Baldauf+16, SteeleBaldauf2020, Chudaykin+21}.
The reason is very simple. We are not trying to determine the domain of validity of the full perturbative expansion (we are actually agnostic regarding its convergence). More pragmatically, we simply want to find out the range of scales for which the one-loop formulae provide an accurate match to 
$N$-body simulations.

In the remainder of this paper,
we distinguish between the concepts of accuracy and validity:
the former indicates how closely
a perturbative expansion reproduces the exact answer
while the latter refers to the consistency of all the assumptions
of the theory. Therefore, the domain of accuracy and the range of validity of the models should not be confused. For instance, a model could still provide a good approximation to the truth on a range of scales although its assumptions are not valid.

\begin{figure*}
 \centering
   \includegraphics[width=\columnwidth]{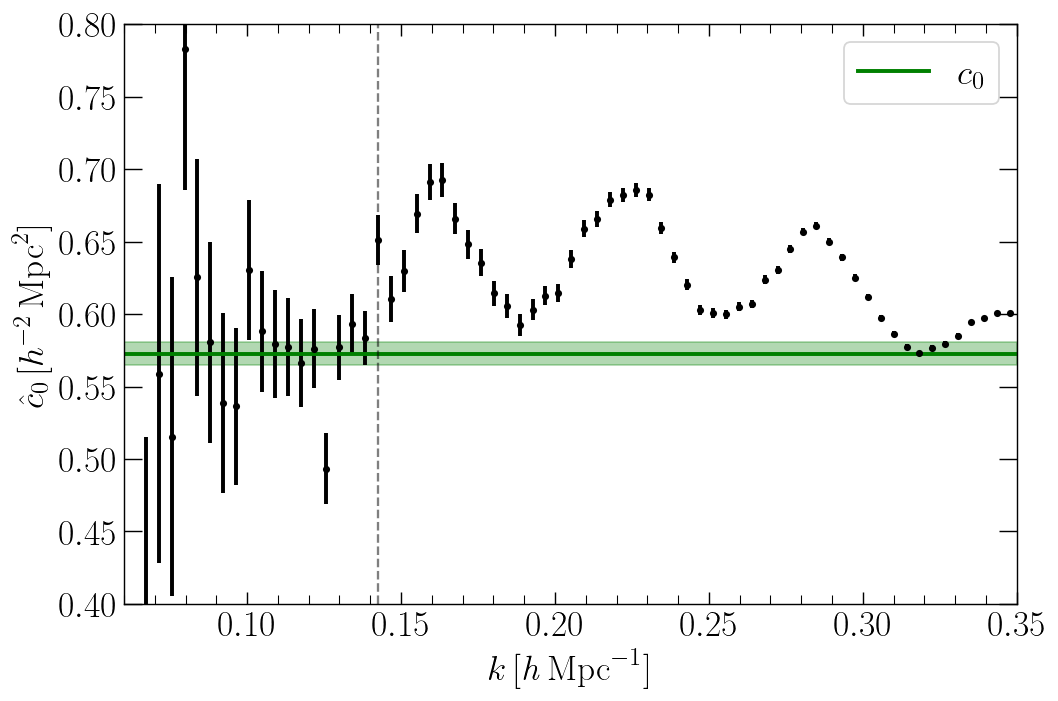}
 \includegraphics[width=\columnwidth]{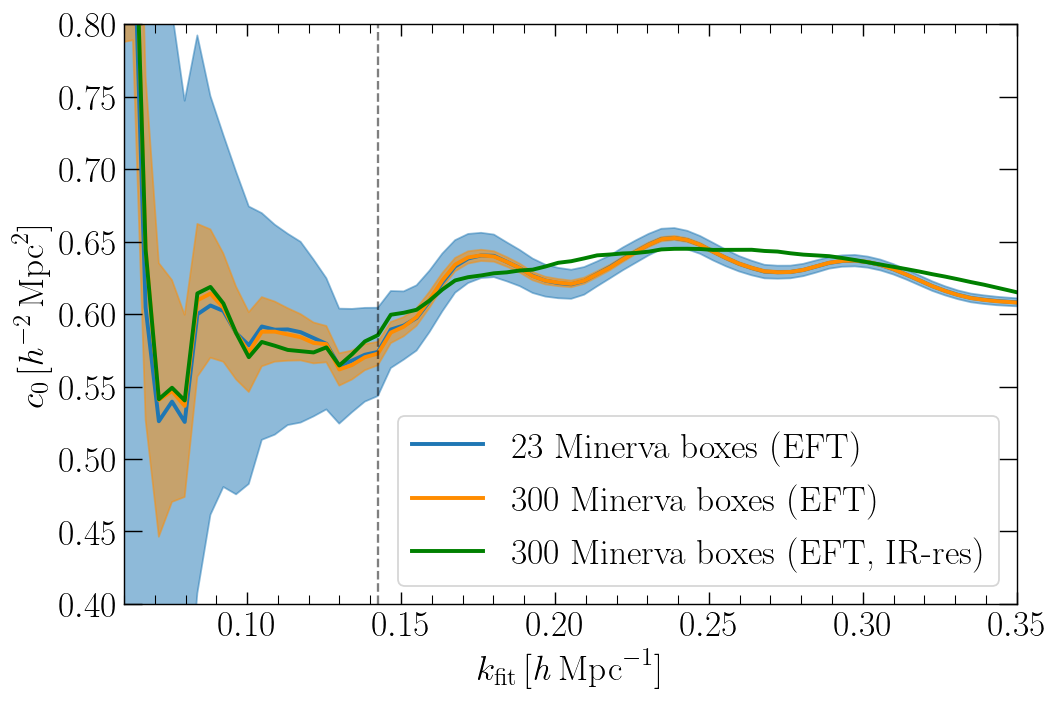}
  \caption{The EFT parameter $c_0$ obtained by matching the model for the power spectrum at one loop to the \minerva simulations. In the left panel, we use the estimator given
  in equation~(\ref{eq:c0esti}),
  while, in the right panel, we fit the 
  model to the numerical data for $k<\kfit$. Shown are the best-fitting
  values (solid lines) and their uncertainty (68-per-cent confidence interval, shaded regions). 
  As indicated in the legend, different colours distinguish results
  obtained with and without accounting for IR resummation or by considering
  subsamples of the \minerva simulations. To improve readability,
  we do not plot the shaded region for
  the IR-resummed EFT case.
  The vertical line indicates the scale at which a statistically significant departure from the low-$k$ limit is detected for the full data set. The horizontal line in the left panel shows the result of the
  fit in the right panel for $\kfit=0.14 \kMpc$.
  }
 \label{fig:eft_ps_fit}
\end{figure*}

\
%%%%%%%%% RESULTS %%%%%%%%%%%%%%%%
\section{Results}
\label{sec:results}

In this section, we determine the domain of accuracy of the perturbative models
for the matter power spectrum and bispectrum we have introduced in section~\ref{sec:theory}
by comparing them against $N$-body simulations.
To start with, we pay
particular attention to discussing
how we fix the EFT parameters that
determine the amplitude of the counterterms.
Subsequently,
we present results as a function
of the total volume used to measure
$P$ and $B$.
As a final step, we discuss the impact
of systematic errors introduced by
the $N$-body technique.

%%%%%%%%%%%%%%%%%%%%%%%%%%%%%%%%%%%%%%%%%%%%%%%%%%
%%%%%%%%%%%%%%%%%%%%%%%%%%%%%%%%%%%%%%%%%%%%%%%%%%

\subsection{EFT parameters}
\label{ssec:fitting_the_eft_params}

As  mentioned in section~\ref{sec:EFT}, the EFT parameters related to the counterterms need to be determined
from the simulation data.
In doing so, we do not distinguish between the actual counterterms and the
renormalised contributions. Therefore,
the coefficients we obtain should be interpreted as simple `matching coefficients' and not given any particular physical interpretation.
Following a common trend in the literature, we will keep referring
to these coefficients as counterterms.

The EFT power spectrum at one loop only contains the parameter $c_0$ for
which, following \citet{Baldauf2015b}, we can build an estimator starting from equations~(\ref{eq:eftP0}) and (\ref{eq:eftP}),
\begin{equation}
\label{eq:c0esti}
    \hat{c}_0 (k)=-\frac{\langle \hat{P}(k)\rangle-P_\text{SPT}(k)}{2\,k^2\, \Pl(k)}\,.
\end{equation}
In the left panel of Fig.~\ref{fig:eft_ps_fit}, we show
how $\hat{c}_0$ depends on $k$ when we use the mean power spectrum extracted from the \minerva simulations.
Within the EFT framework, $c_0$ is a scale-independent parameter but our data
show that $\hat{c}_0$ significantly deviates from its low-$k$ limit
when $k>0.14 \kMpc$. 
This is usually interpreted 
as a signal that
the truncated perturbative expansion breaks down beyond this scale and higher-order corrections become important \citep{Foreman-Perrier-Senatore-2016}. 
In the right panel of Fig.~\ref{fig:eft_ps_fit},
we determine $c_0$ by fitting $P_\mathrm{EFT}$ (with and without IR resummation) to the mean power spectrum extracted from the \minerva simulations. Our results are shown as a function of the maximum wavenumber used in the fit, $\kfit$.
The orange line represents the best-fitting value for the EFT model and the shaded region around it
marks the 68-per-cent confidence region
of the fit.
Not surprisingly, it resembles a smoothed version 
of the results shown in the left panel.
Considering subsets of 23 \minerva boxes (which cover the same total volume as the \textsc{Eos} simulations) only increases the scatter of the estimates
(blue shaded region).
Accounting for the IR resummation (green line) removes the oscillations in the
region of the baryonic acoustic features
but does not attenuate the overall
scale dependence for $k>0.14 \kMpc$.
Based on these results,
we conclude that the domain of validity of the one-loop EFT expressions for the power spectrum
at $z=1$ is $k<0.14 \kMpc$.
Our results are consistent with fig.~14 of \citet{Baldauf2015b}, even though our analysis is performed at $z=1$ instead of $z=0$. Remarkably, the limiting value we find is also consistent with the blinded challenge presented in \citet{Nishimichi2020}, which uses a total simulation volume of 566$\cGpc$ (about half of the volume covered by the \minerva suite) at $z=0.61$ to test
the constraining power for cosmology of
the EFT predictions for the galaxy
power spectrum in redshift space.
In this case, the recovered cosmological parameters show a bias whenever the mock data sets are extended beyond $\kmax = 0.14\,\kMpc$.
In the remainder of this paper,
we use the best-fitting value of $c_0$
using $\kfit = 0.14\,\kMpc$ as the default
option for $P_\mathrm{EFT}$. This gives
$c_0 = 0.581\pm 0.009\,h^{-2}\,\mathrm{Mpc}^2$.
If we simply rescale this value by $[D(z=1)]^{-2}$ (thus ignoring any intrinsic time dependence of $c_0$), we obtain 1.525 which closely matches the results previously obtained at $z=0$ using slightly
different cosmological models, methods, and scales \citep{Carrasco2014b, Angulo2015, Baldauf2015}.

\begin{figure*}
 \centering
 \includegraphics[width=\textwidth]{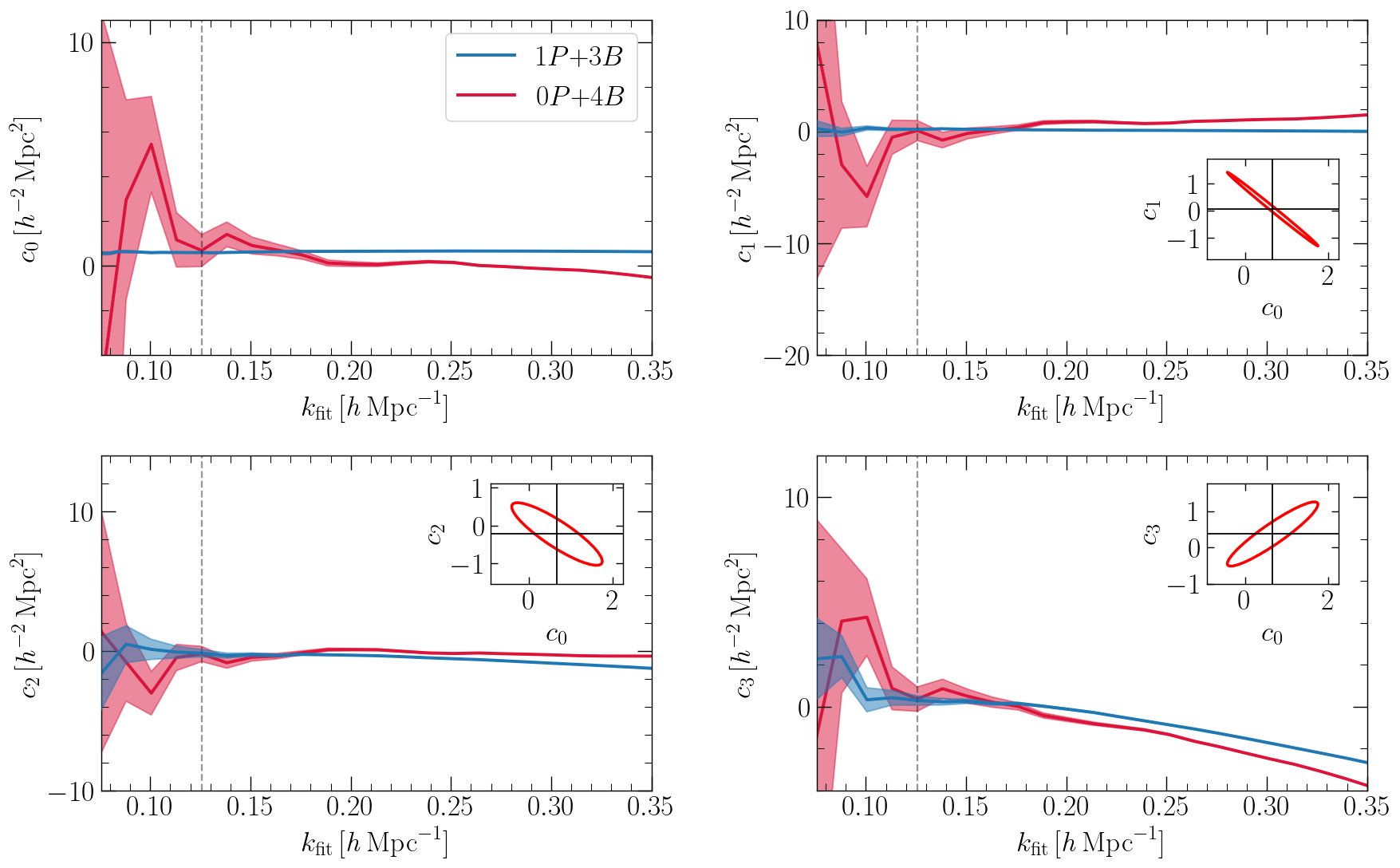}
  \caption{The best-fitting values (solid lines) and the uncertainties (shaded regions) of the EFT parameters that influence the matter bispectrum at one loop are shown as a function of the maximum wavenumber of the measurements extracted from the \minerva simulations.
  We consider two cases: (i) we fix $c_0$ using the power spectrum (Fig.~\ref{fig:eft_ps_fit}) and the other three parameters using the bispectrum (blue) and (ii) we fit all four parameters using the bispectrum measurements (red).
The vertical dashed lines indicate
the default value of $\kfit=0.125\,\kMpc$ we use in the remainder
of the paper.
The insets show the joint 68-per-cent confidence regions for two EFT parameters evaluated for $\kfit=0.125\,\kMpc$. The cross indicates the best-fitting values of these parameters.}
 \label{fig:eft_bisp_fit}
\end{figure*}

The EFT bispectrum at one loop contains 
four unknown parameters and 
different strategies have been 
developed in the literature to determine them. For instance, it is possible
to express $c_1, c_2$ and $c_3$ as a function of $c_0$
by imposing renormalisation conditions
and then 
fix $c_0$ from a fit to  the power spectrum
\citep{Angulo2015, Baldauf2015}.
An alternative line of attack -- which we follow here -- is to treat (at least some of) the EFT parameters as fit parameters for the bispectrum,
trying to avoid overfitting.
The recent and comprehensive study by \citet{SteeleBaldauf2020} gives evidence supporting the second approach. 
Two options are available when fitting
$c_0$: we can either use the value that best fits the power spectrum or determine it together with the other EFT parameters only using the bispectrum.
We compare these alternatives in Fig.~\ref{fig:eft_bisp_fit}, where we
show the dependence of the best-fitting EFT parameters on the maximum wavenumber
considered in the fit. In all cases, we use the IR resummed model.
Focussing on $c_0$, we notice that
the fit based on the power spectrum
is much more stable and less uncertain at large scales. 
It is also worth stressing that the best-fitting values are sometimes negative while $c_0\geq 0$ in the theory. However, we do not give much weight to this consideration since we do not fit the renormalised counterterms.
We also notice that $c_0$ and the
other EFT parameters
are strongly correlated when they are simultaneously fit from the bispectrum.
Their trend with $\kfit$ in
Fig.~\ref{fig:eft_bisp_fit} clearly shows that this happens at all scales. Such degeneracy is investigated in more detail in the three insets 
where we show the joint
68.3-per-cent confidence region for
$c_0$ and a second parameter estimated at $\kfit= 0.125\,\kMpc$ while keeping the remaining two fixed at their best-fitting values.
The cross-correlation coefficients
between $c_0$ and the other parameters are as high as $-0.997$ ($c_1$), $-0.872$ ($c_2$) and $0.922$ ($c_3$) 
indicating that the bispectrum data
cannot isolate the contributions
from the different counterterms.
We believe that the large-scale fluctuations of the EFT parameters
are largely influenced by this degeneracy. The fluctuations are, in fact, greatly suppressed by fixing $c_0$ to the best-fitting value from the power spectrum. In this case, the four EFT parameters assume consistent values for
 $\kfit<0.14 \kMpc$ (which is basically determined by $P$). On the contrary, considering
 smaller scales generates deviations,
 especially for $c_0$ and $c_3$.
 All this suggests that $k= 0.14 \kMpc$ also approximately delimits the domain of validity of the one-loop model
 for the bispectrum at $z=1$.
However, the two fitting methods (i.e. fitting three or four counterterms with the bispectrum)
do not provide fully consistent results for all parameters at $\kfit\approx 0.14\kMpc$ while they do at $\kfit\approx 0.125\kMpc$ (the actual values are listed in Table~\ref{tab:counterfit}). For this reason,
unless explicitly stated otherwise, from now on we fix the EFT parameters for $B_\mathrm{EFT}$ to the best-fitting values at this scale
using the power spectrum to determine
$c_0$ and the bispectrum to measure $c_1, c_2$ and $c_3$.
Note that, in spite of our unprecedentedly large data set, only $c_0$ is precisely determined
while $c_1, c_2$ and $c_3$ are compatible with being zero within a few standard deviations.
This suggests that it might be quite challenging to fix the EFT counterterms for the bispectrum from actual observational data and all what could be done is to marginalise over them in a Bayesian fashion, possibly reducing the
constraining power for cosmology of the data.
\begin{table*}
	\centering
	\bgroup
	\def\arraystretch{1}
	\begin{tabular}{c|c|c|c|c|c}
		\hline
	Fit& $\chim$ &	$c_0$ & $c_1$ & $c_2$ & $c_3$\\
	& &	$[h^{-2}\,\mathrm{Mpc}^2]$&
		$[h^{-2}\,\mathrm{Mpc}^2]$&
		$[h^{-2}\,\mathrm{Mpc}^2]$&
		$[h^{-2}\,\mathrm{Mpc}^2]$\\
		\hline
	$1P+3B$& $171.03/167$	& $0.577 \pm 0.013$ 
		& $0.177 \pm 0.071$ 
		& $-0.16 \pm 0.27$ 
		& $0.30 \pm 0.23$ \\ 
	$0P+4B$ & $171.01/166$		& $0.67 \pm 0.72$ 
		& $0.06 \pm 0.90$ 
		& $-0.22 \pm 0.54$ 
		& $0.37 \pm 0.58$ \\ 	
		\hline
	\end{tabular}
	\egroup
	\caption{Best-fitting values and uncertainties
	of the EFT parameters derived with $\kfit= 0.125 \kMpc$ in Fig.~\ref{fig:eft_bisp_fit}. The $\chim$ statistic for the best fit is expressed in terms of $\chi^2_\mathrm{tot}$ and $\nu$ as in equation~(\ref{eq:chi_2}).}
	\label{tab:counterfit}
\end{table*}

\begin{figure}
 \centering
 \includegraphics[width=\columnwidth]{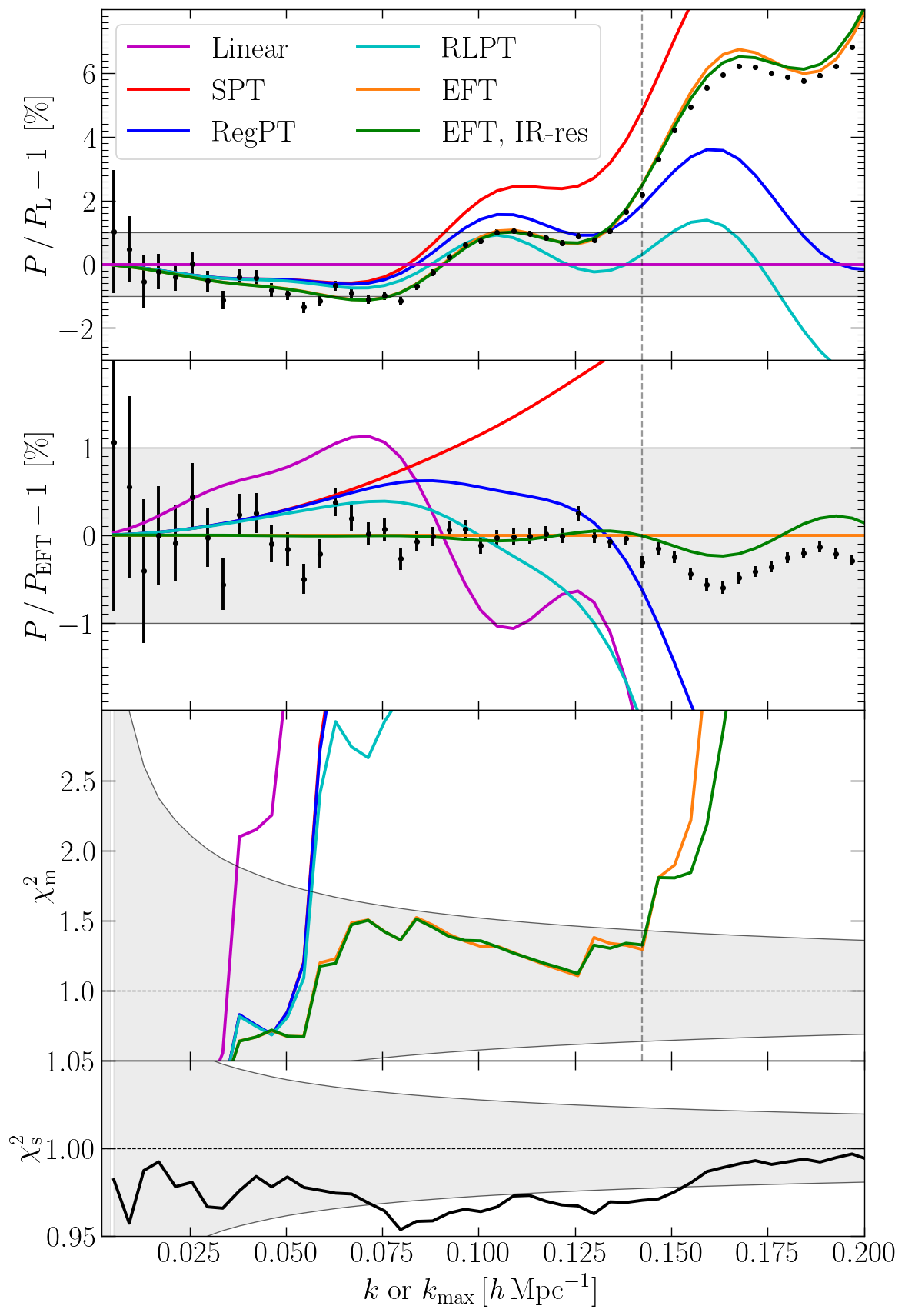}
 \caption{Comparison of the perturbative
 models for the matter power spectrum
 (solid lines)
 with the mean measurement
 extracted from the \minerva simulations (symbols with error bars) using the bin width $\Delta k = \kF$. The top two panels compare the different power spectra to $P_\mathrm{L}$ and
 $P_\mathrm{EFT}$ whereas the bottom two
display the goodness of fit for the models and the covariance matrix (see section~\ref{ssec:validation_of_models} for details). 
The grey shaded areas in the top two panels represent deviations smaller than
one per cent
with respect to the reference models.
Those in the bottom two panels, instead,
mark the regions bounded by the upper and lower 95 per cent confidence limits for the $\chi^2$ distribution with the appropriate
number of degrees of freedom. The vertical dotted line indicates $\kfit=0.14 \kMpc$ which is the largest wavenumber used to fit $c_0$ for the EFT models.}
 \label{fig:ps_chi_s1}
\end{figure}
\subsection{Power spectrum}
\label{sec:power_spectrum}

If we assume for the moment that
the $N$-body technique does not introduce any systematic shifts, 
we can determine the domain of accuracy of the perturbative models for the matter power spectrum by directly comparing
their predictions to the measurements
extracted from the numerical simulations.
The huge volume covered by the \minerva simulations results in sub-per-cent
statistical uncertainties for the
average power spectrum (see Fig.~\ref{fig:ps_numerical}).
Getting agreement to this precision
would be a major achievement for the models.

In the top panel of Fig.~\ref{fig:ps_chi_s1},
we show the mean power spectrum
extracted from the \minerva simulations
using $\Delta k = \kF$ (symbols with error bars) and the corresponding bin-averaged models (solid lines with different colours as indicated by the label). In order to reduce the span of the data and improve readability, we plot
the deviation from the linear power spectrum in per-cent points. Similarly,
in the second panel from the top, 
we show the same measurements and models
but in terms of their deviation with respect to $P_\mathrm{EFT}$ which provides a better fit to the numerical data. The third panel, instead,
shows the $\chim$ statistic evaluated for the different models as a function of $\kmax$. This quantity gives a measure of the goodness of fit. In order to have a reference scale, we highlight
the regions bounded by the one-sided upper and lower 95 per cent confidence limits for the $\chi^2$ statistic (with the appropriate number of degrees of freedom)
with a grey shaded region. 
Basically, a model should be rejected
at 95 per cent confidence when its $\chim$ lies outside the shaded region.
In practice, we determine the domain of accuracy of the models as follows: moving from left to right,
we look for the first $\kmax$ at which
$\chim$ lies outside the shaded region.
Finally, the bottom panel shows the $\chis$ statistic as a function of $\kmax$ together with the corresponding
95 per cent confidence limits
that can be used to
evaluate the quality of our approximation
for the covariance matrix.

Coming to the specific outcome of this comparison, Fig.~\ref{fig:ps_chi_s1} indicates that, although equation (\ref{eq:Cpp})
systematically overestimates the variance
of our measurements by a few percent, this discrepancy is hardly statistically significant in the range of scales we consider.
Moreover, 
by examining the $\chim$ curves
as a function of $\kmax$, it is evident that, when a model begins to break down, $\chim$ increases very steeply so that the inferred reach is quite insensitive to 
small deviations in the size of the error bars.  
Therefore, we can proceed further with
analysing the $\chim$ curves knowing that this statistic will be only very slightly underestimated. This provides
a clear ranking for the models based
on their domain of accuracy.
Not surprisingly, the first model to break down is $\Pl$ which fits the
\minerva simulations only for
$\kmax<0.035 \kMpc$, followed by SPT
($\kmax < 0.06\kMpc$).
Since on these relatively large scales
RegPT and RLPT essentially coincide with SPT, they also fail at the same $\kmax$. Contrary to SPT, however, they agree
with the simulations to better than 
one per cent up to $k\simeq 0.15\kMpc$.
The best agreement is found with the EFT model which fits the \minerva simulations accurately for $k_\mathrm{max}< 0.14\kMpc$ (and never shows per cent deviations within the explored range of wavenumbers). Consistently with previous work \citep[e.g.][]{Baldauf-ir-res2015}, we find that IR-resummation improves the fit only beyond its nominal range of accuracy. 
One issue worth investigating is that the value of $\chim$ rises sharply around $k\approx0.125\,\kMpc$ and $0.14\,\kMpc$, which causes the EFT models to get rejected on slightly larger scales than perhaps expected (based on visual inspection of the top panel in Fig.~\ref{fig:ps_chi_s1}). This is caused by the statistically significant deviation of two simulation data points around those scales which are clearly distinguishable in the second panel (from the top) of Fig.~\ref{fig:ps_chi_s1}. After carefully inspecting individual
simulations to understand the origin of these deviations, we could not reach any clear conclusion. However, upon re-measuring the power spectrum using narrower bins in that region, we notice that the deviations form coherent features within a range of $k$-values and are not simply due to random noise.

\begin{figure}
 \centering
 \includegraphics[width=\columnwidth]{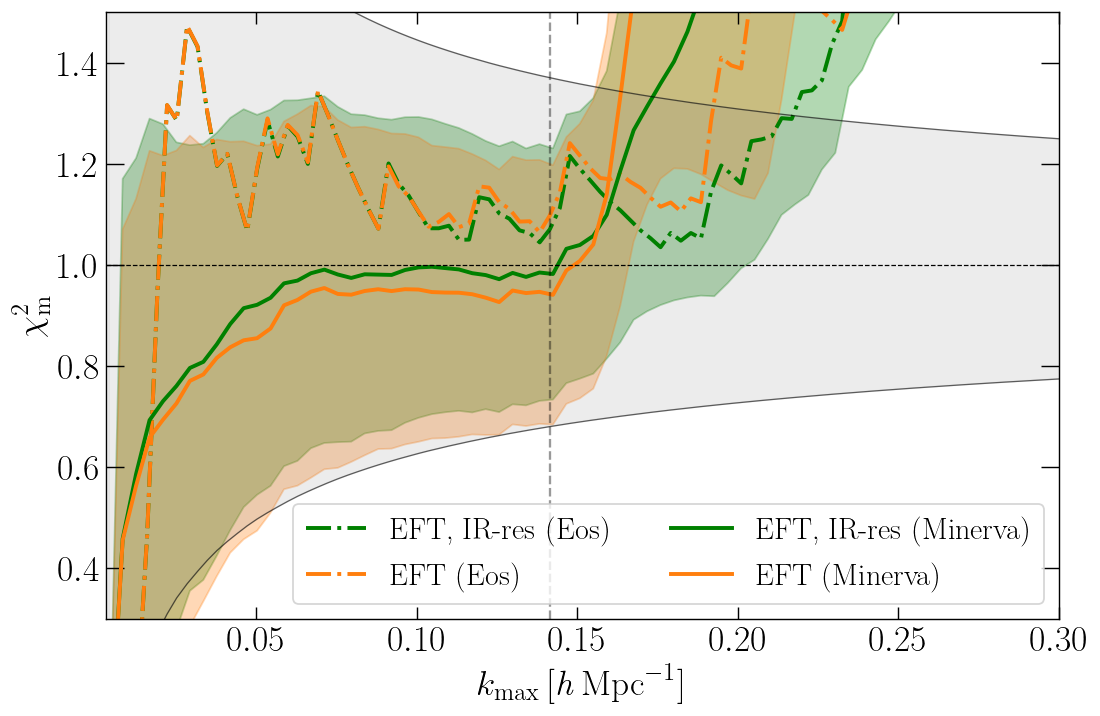}
 \caption{The $\chim$ statistic 
 for the EFT and \eftir{} power spectra
 evaluated using the \textsc{Eos} suite (dash-dotted) and many different subsets of 23 \minerva simulations (solid and shaded for mean and standard deviation, respectively). 
The grey shaded area highlights the region bounded by the (one-sided) upper and lower 95-per-cent confidence limits for a $\chi^2$ distribution function with the appropriate number of degrees of freedom. The vertical dotted line indicates $\kfit$.}
 \label{fig:ps_chi_s1_eos_minerva}
\end{figure}
\begin{figure}
 \centering
 \includegraphics[width=\columnwidth]{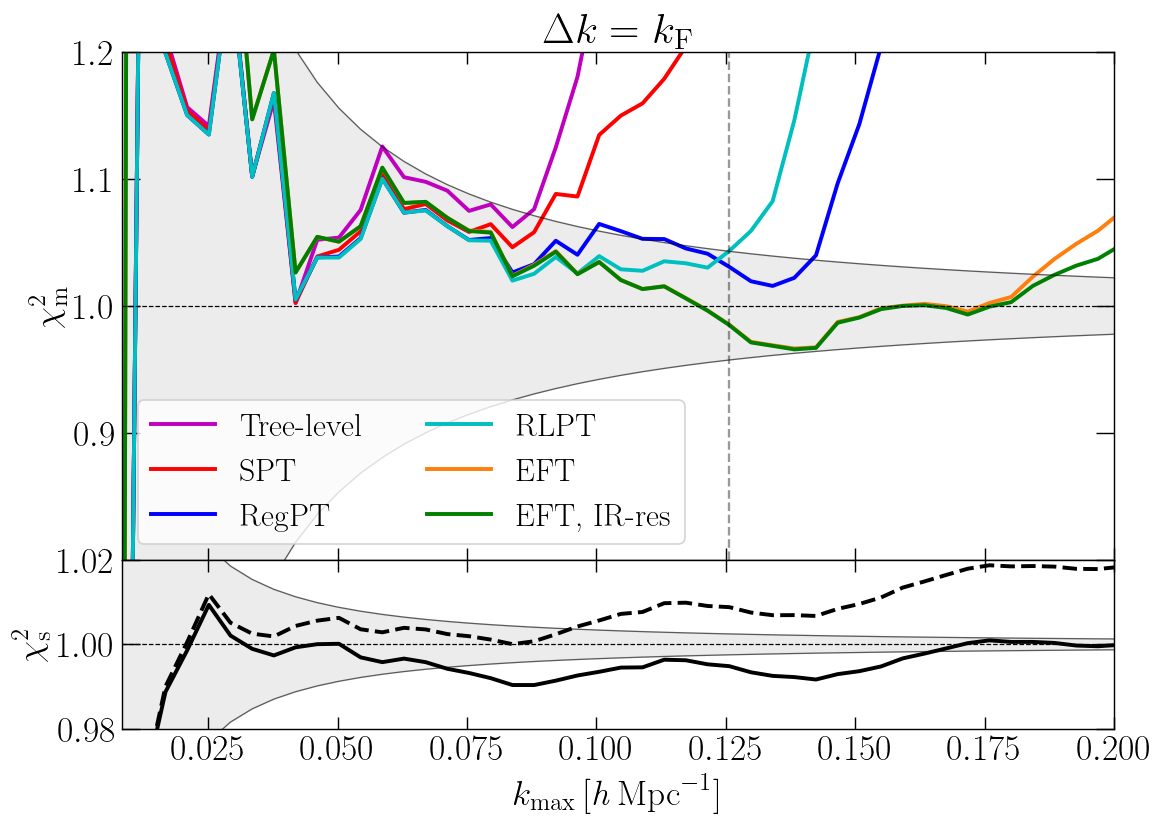}
 \includegraphics[width=\columnwidth]{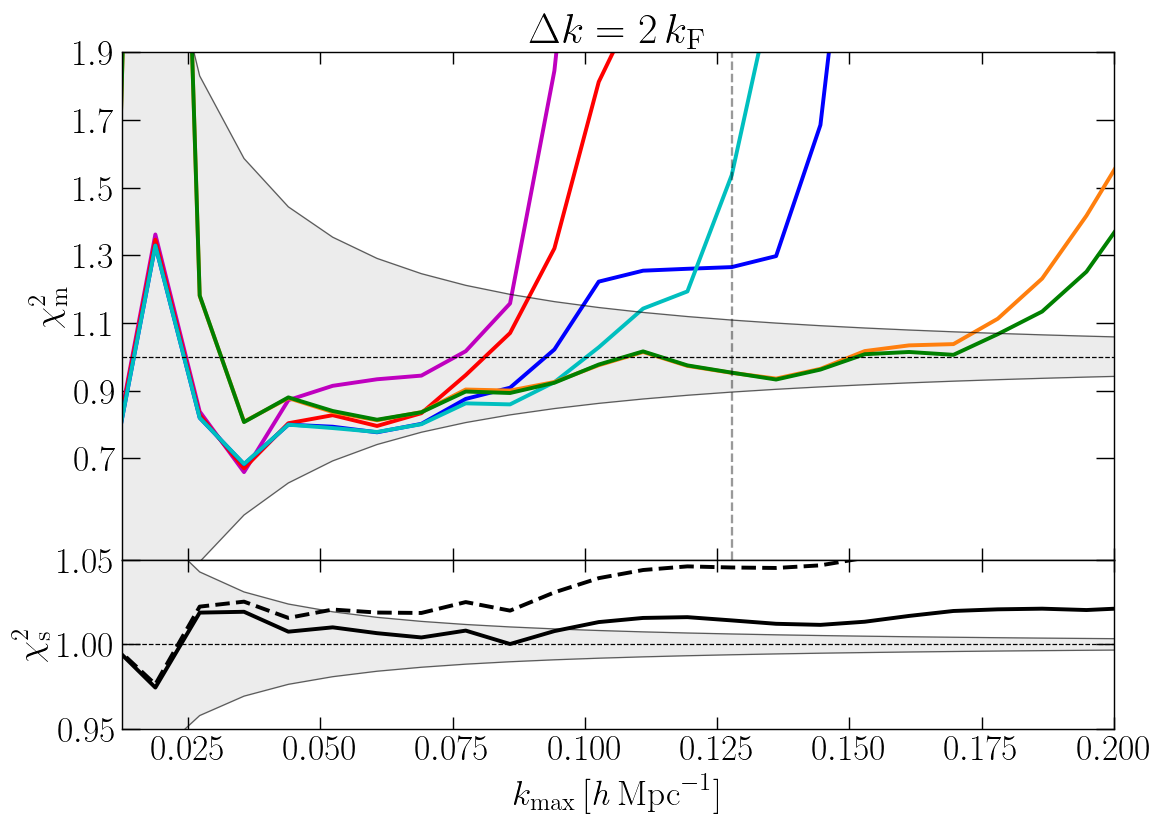}
 \includegraphics[width=\columnwidth]{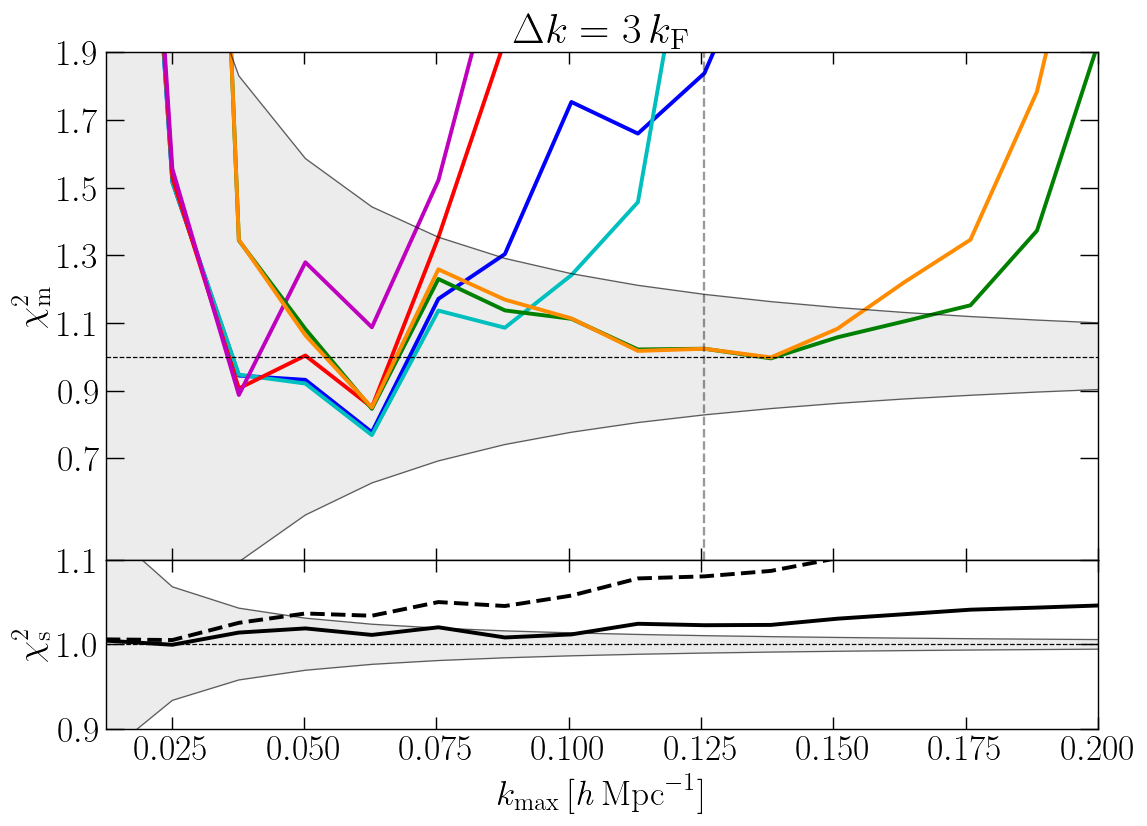}
 \caption{As in the bottom two panels of Fig.~\ref{fig:ps_chi_s1} but for the matter bispectrum. From top to bottom, we consider
 three different bin widths, namely
 $\kF$, $2\,\kF$ and $3\,\kF$. The dashed lines in the plots for $\chis$ (bottom panels) refer to the Gaussian approximation to the covariance matrix.} 
 \label{fig:bs_chi_s1}
\end{figure}

To cross check our results and also test the models under less demanding standards, we repeat our analysis
using the \textsc{Eos} simulations
which cover a smaller volume (roughly corresponding to 23 \minerva boxes) and thus give larger statistical error bars.
For simplicity, we only consider the EFT models with and without IR-resummation. 
In order to properly compare results obtained using the \minerva and \textsc{Eos} suites, we proceed as follow: (i) we random sample 23 \minerva boxes from the full set; (ii) we fit the EFT parameter $c_0$ to the mean power spectrum of the subset using $\kmax=0.14 \kMpc$; (iii) we compute the $\chim$ statistic as a function of $\kmax$ for the best-fitting $c_0$.
Our results are shown in Fig.~\ref{fig:ps_chi_s1_eos_minerva} where 
the solid lines represent the mean
$\chim$ obtained from the \minerva subsets and the shaded regions around them show the corresponding standard deviation.
Overall, these findings are in very good agreement with the $\chim$ curves derived from the 
\textsc{Eos} simulations (dot-dashed lines).
Due to the larger statistical error bars, the nominal reach of the EFT models slightly increases with respect to the analysis performed with the full
\minerva set. We find 
$\kmax < 0.16_{-0.01}^{+0.05} \kMpc$ for standard EFT
and $\kmax< 0.17_{-0.02}^{+0.06} \kMpc$ for IR-resummed EFT. Note that this extends beyond the minimum scale for which $c_0$ can be assumed to be constant.

\subsection{Bispectrum}
\label{sec:bisp}

\begin{figure}
 \centering
 \includegraphics[width=\columnwidth]{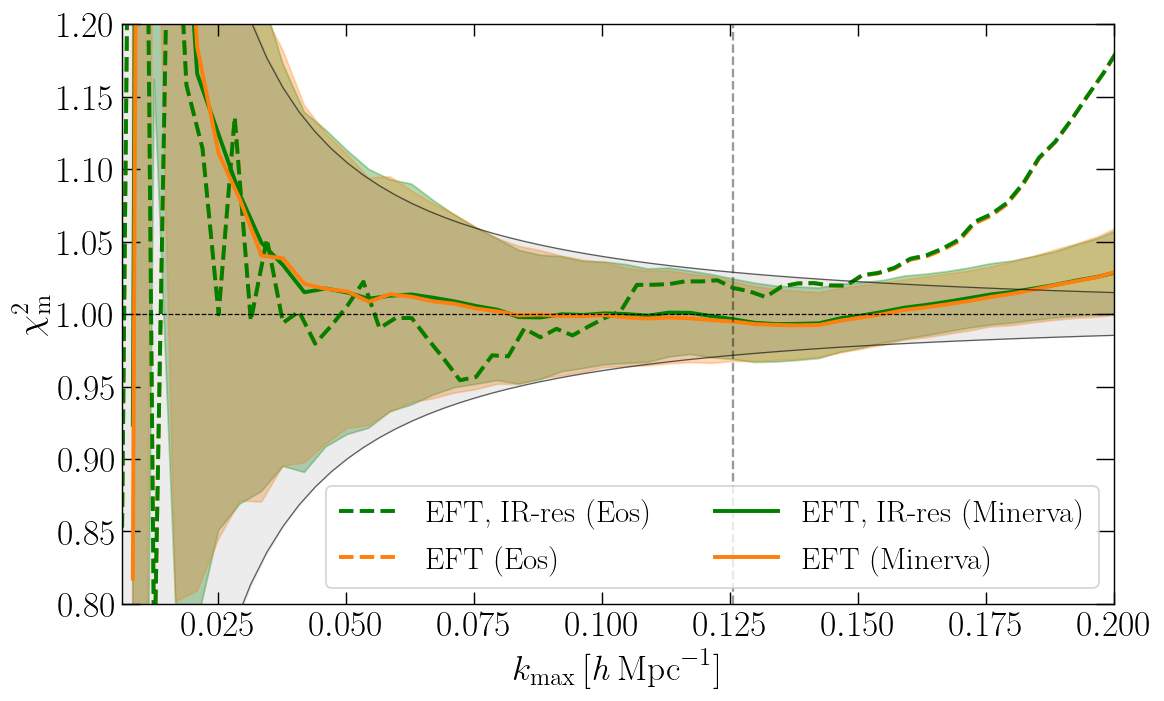}
 \caption{As in Fig.~\ref{fig:ps_chi_s1_eos_minerva} but for the matter bispectrum.}
 \label{fig:bisp_chi_s1_eos_minerva}
\end{figure}

In Fig.~\ref{fig:bs_chi_s1}, we investigate the goodness of fit of the different models for the matter bispectrum
by plotting the $\chim$ and $\chis$ statistics as a function of $\kmax$
for various bin sizes. 
In the bottom panels, we show two curves: the solid one considers our approximation to the covariance matrix
given in equation~(\ref{eq:C_BB})
while the dashed one refers to
the Gaussian part given in
equation~(\ref{eq:C_BBGauss}).
It is evident that the Gaussian approximation severely underestimates the variance of the bispectrum measurements already at large scales and especially for broader triangle bins. On the contrary, equation~(\ref{eq:C_BB})
provides average deviations of only a few per cent for all configurations considered in this work. We believe that this is accurate enough to get robust estimates of $\chim$, although, at small scales, the assumed covariance matrix
is nominally incompatible with the scatter seen in the simulations (i.e.
the black curve lies outside the shaded region in the plots for $\chis$).

The variations of $\chim$ with $\kmax$
provide a clear ranking of the models,
Independently of the bin width,
the tree-level SPT prediction 
breaks down first 
and one-loop corrections only slightly
improve the range of accuracy of the theory up to $k_\mathrm{max}\simeq 0.08 \kMpc$.
RegPT and RLPT provide substantial improvements and accurately match the \minerva simulations 
up to scales between $0.1\kMpc$ and $0.14\kMpc$ depending on the bin width.
Finally, the counterterms in
the EFT bispectra boost the agreement up to $k_\mathrm{max}\simeq 0.16-0.19\kMpc$. IR-resummation turns out to be relevant only for $k\gtrsim 0.15\kMpc$ and even marginally for the case of narrow bins, where statistical errors are larger than the deviations between the model and the data at the scales of the baryonic acoustic oscillations.

As we already did with the power spectrum, in Fig.~\ref{fig:bisp_chi_s1_eos_minerva},
we verify that using
the \minerva and \textsc{Eos} simulations gives consistent results
for the bispectrum as well.
It turns out that
the reach of the EFT models is a bit reduced for the \textsc{Eos} simulations but this is consistent with random fluctuations.
We also note that, for \textsc{Eos}, the EFT models with and without IR-resummation have practically the same domain of accuracy as a consequence of the larger uncertainty of the measurements.

\begin{figure}
 \centering
 \includegraphics[width=\columnwidth]{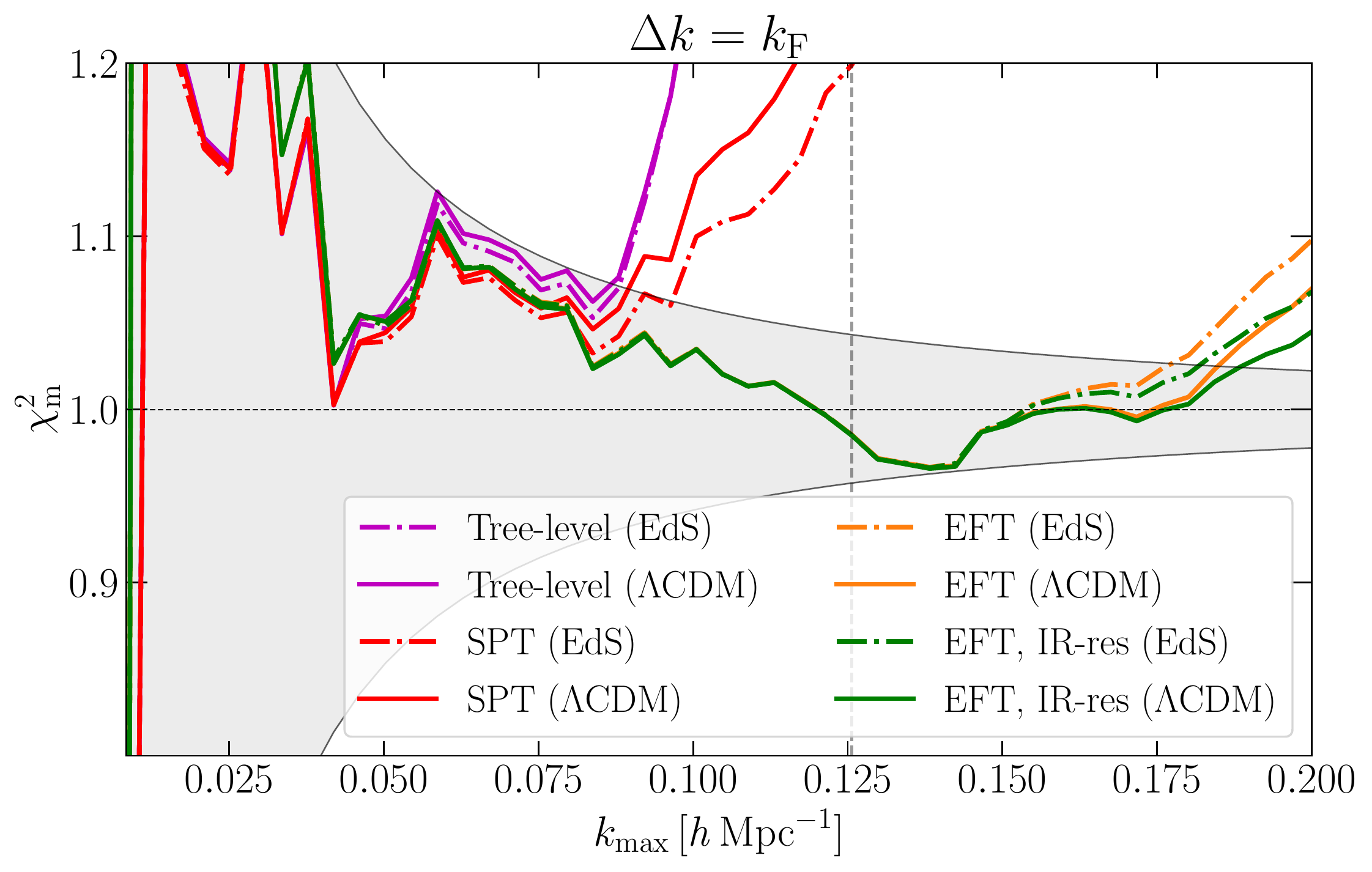}
 \caption{As in the top panel of Fig.~\ref{fig:bs_chi_s1} but comparing models obtained with
 the exact second-order SPT kernels 
 ($\Lambda$CDM) and the popular EdS approximation (see the main text for details).}
 \label{fig:bisp_chi_s1_lcdm_vs_eds}
\end{figure}

Finally, we test the impact of using
the popular EdS approximation for the second-order kernel $F_2$ instead of the more general scheme we described in section~\ref{sec:timeevo}.
Fig.~\ref{fig:bisp_chi_s1_lcdm_vs_eds}
shows that this modification has very little influence on our results.
No changes in the $\chim$ are visible on large scales where error bars are bigger. The only noticeable differences are: (i) a slight improvement in the reach of SPT when the EdS approximation is adopted and (ii) a similarly sized increase in the range of accuracy of EFT when the exact second-order kernel is used.

%%%%%%%%%%%%%%%%%%%%%%%%%%%%%%%%%%%%%%%%%%%%%%%%%%%%%
\subsection{Range of accuracy vs surveyed volume}
\label{sec:systematic_errors}
\begin{figure*}
	\includegraphics[width=\textwidth]{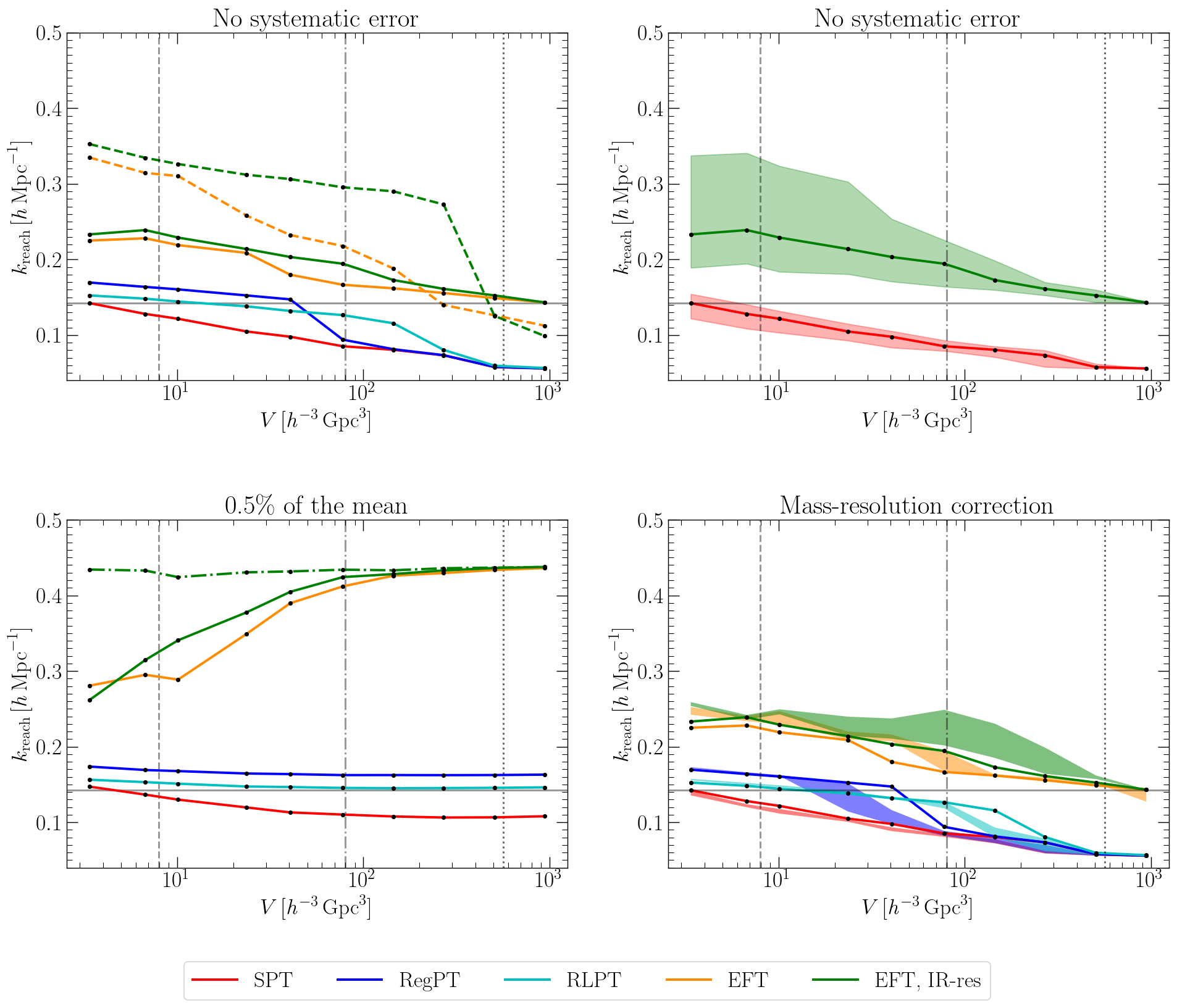}
	\caption{We define the reach of a model for the power spectrum as the minimum $\kmax$
	at which the $\chi^2$ goodness-of-fit test rejects the null hypothesis that
	the $N$-body data are consistent with the model predictions at the significance level of 0.05. The top-left panel shows the median reach of 200 subsets of \minerva simulations each covering a volume $V$. Different colours refer to different models as indicated in the label. Solid lines are used for the models with no free parameters and
	for our default EFT models (i.e. with  $\kfit=0.14\kMpc$, highlighted by a horizontal grey line) while the two dashed lines represent the EFT models with $\kfit=0.22\kMpc$.
	The top-right panel shows the median (solid) and the central 68-per-cent range (shaded) of the estimated reach for SPT and the default \eftir{}.
	The bottom-left panel is analogous to the top-left one but accounts for systematic errors in the simulations by considering an additional 0.5-per-cent error added in quadrature to the random contributions. The dot-dashed line refers to the \eftir{} model obtained by averaging $c_0$ over the 200 subsets. Finally, the bottom-right panel shows the reach of the models after approximately correcting the simulation data for
	the bias introduced by the finite mass resolution (see the main text for details). The shaded regions encompass the range of variability of the corrections
	while 
	the solid lines are taken from the top-left panel and are given as a reference. 
	All panels show three vertical lines indicating: (i) the volume of a redshift bin of width $\Delta z=0.2$ centred at $z=1$ for a Euclid-like survey (dashed); (ii) the total volume of the \textsc{Eos} simulations (dot-dashed); (iii) the volume of the PT-challenge simulations in 
	\citet[][dotted]{Nishimichi2020}.
	Measurements and models are compared using a bin width of $\Delta k = \kF$.
	}
	\label{fig:p_k_vs_V_s1}
\end{figure*}
\begin{figure*} 
	\centering
	\includegraphics[width=\textwidth]{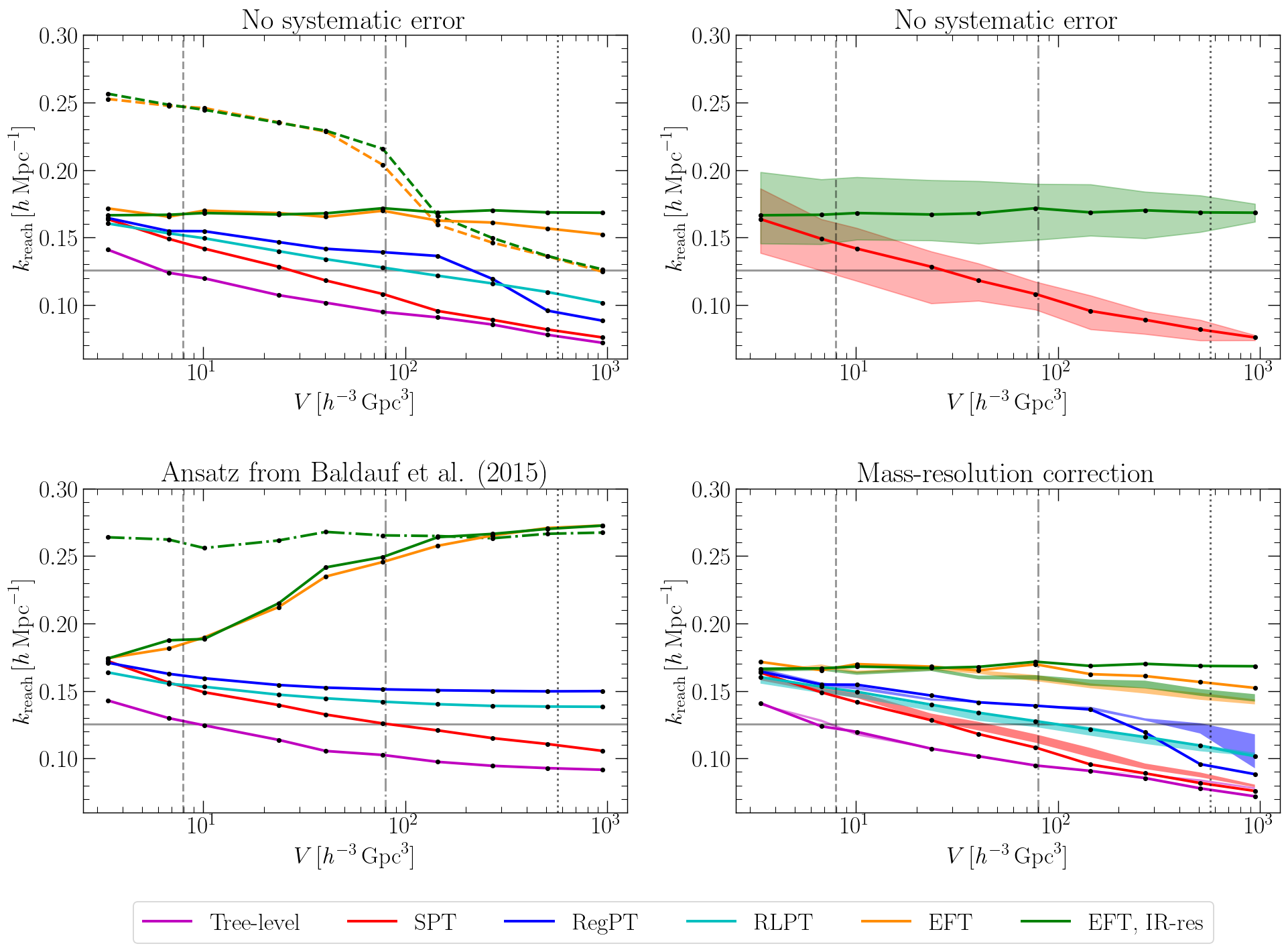}
	\caption{As in Fig.~\ref{fig:p_k_vs_V_s1} but for the bispectrum models and using $\Delta k = 3\,\kF$. In the bottom-left panel,
	 systematic errors were added in quadrature to the statistical errors of the measurements
	 following the ansatz of \citet{Baldauf2015}.}
	\label{fig:fig_bisp_k_vs_V_s3}
\end{figure*}
\begin{figure*}
 \centering
 \includegraphics[width=\textwidth]{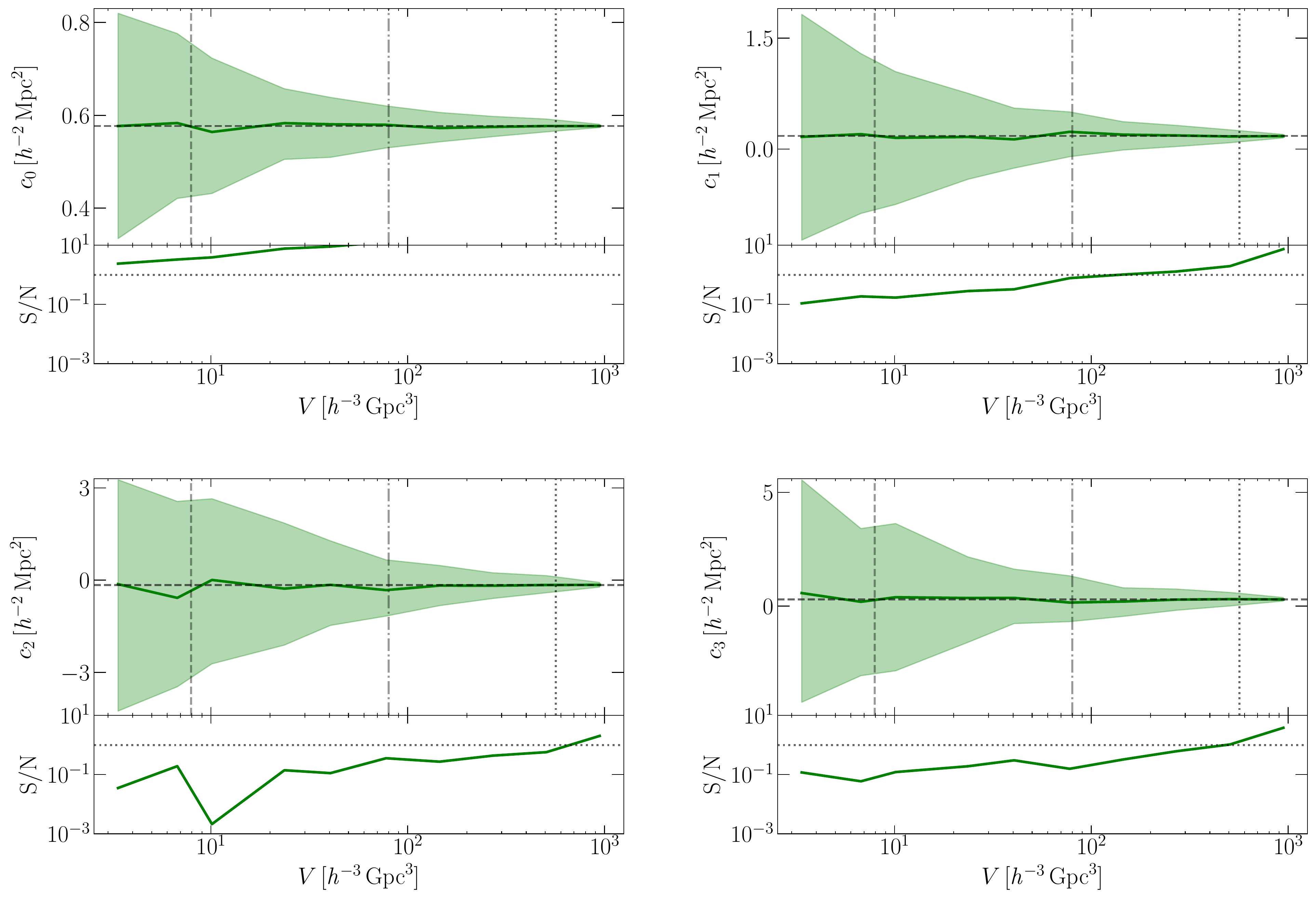}
 \caption{Distribution of the best-fitting EFT parameters 
 as a function of the volume $V$ used to measure the power spectrum and the bispectrum.
The data refer to the counterterms of the \eftir{} model with $\kfit=0.125 \kMpc$. The parameter $c_0$ is derived
from the power spectrum while $c_1, c_2$ and $c_3$ are determined from the bispectrum.
 The top panels show the median (solid) and the central 68-per-cent region (shaded) over 200 realisations. The horizontal dashed lines indicate the values obtained from the full \minerva simulation suite and given in Table~\ref{tab:counterfit}. The three vertical lines mark the same volumes as in Figs.~\ref{fig:p_k_vs_V_s1} and \ref{fig:fig_bisp_k_vs_V_s3}.
 The bottom panels show the signal-to-noise ratio defined as the ratio between the median and 
 and half the central 68-per-cent interval.}
 \label{fig:countervsV}
\end{figure*}

The careful reader might have noticed
that our results are more conservative than other estimates in the literature.
This is partially due to the fact that we use all bispectrum configurations, but mostly because we use a much larger
set of $N$-body simulations
\citep[cf.][]{Angulo2015}.
Current surveys of the large-scale structure of the Universe cover volumes which are one to two orders of magnitude smaller than the total volume of the \minerva simulations.
This directly translates into
larger statistical uncertainties
for summary statistics like the power spectrum and the bispectrum
and thus into more extended ranges of accuracy for the models.
In this section, we investigate
how the reach of the models
depends on the volume covered by a survey. In doing so, we also need to account for systematic effects which we have so far neglected.
In order to better evaluate their impact on our conclusions, we start with assuming that they are of no consequence (we relax this assumption in the next section).

In the top panels of Figs.~\ref{fig:p_k_vs_V_s1} 
and \ref{fig:fig_bisp_k_vs_V_s3},
we show
how the reach of models for $P$ and $B$ changes with the volume
over which the measurements are performed.
These plots are obtained as follows.
(i) We pick a volume $V$ which corresponds to an integer number $N$ of \minerva boxes.
(ii) We randomly select $N$
\minerva realisations (with no repetitions)
and compute
$\langle \hat{P} \rangle$ and $\langle \hat{B} \rangle$ (using $\Delta k=\kF$ for $P$ and $3\,\kF$ for $B$ in order
to probe a wider range of scales).
(iii) We fit the EFT counterterms to the numerical data using $\kfit=0.14 \kMpc$ for $P$ and $\kfit=0.125 \kMpc$ for $B$.
(iv) We evaluate the $\chim$ statistic as a function of $\kmax$ and use it to determine the reach of each model based on the (one-sided) 95-per-cent confidence limits for the chi-squared distribution.
(v) We repeat the procedure from step (ii) onward 200 times.
(vi) We plot the median value of the reach (top-left panels) and
its scatter (top-right panels) as a function of $V$.

In order to ease the interpretation
of our results and facilitate comparison with the literature, we draw vertical
lines marking three characteristic volumes. From left to right, they are:
(i) $V=7.94 \cGpc$ which corresponds to a redshift bin centred at $z=1$ and of width $\Delta z =0.2$ of a Euclid-like survey \citep{EuclidForecast2019};
(ii) $V=80\cGpc$ which coincides with the total volume of the \textsc{Eos} simulations \citep[and is approximately a factor 1.5 larger than the volume of the simulations used in][]{Baldauf2015b,Baldauf2015,SteeleBaldauf2020}; (iii) $V=566\cGpc$ which
is the volume of the simulation used in the blinded challenge paper of \citet{Nishimichi2020}.

We are now ready to discuss the results
presented in the top-left panels of Figs.~\ref{fig:p_k_vs_V_s1} 
and \ref{fig:fig_bisp_k_vs_V_s3}.
As expected, the domain of accuracy of the models decreases with increasing $V$. The only exception is the case
of the EFT bispectrum for which the
reach turns out to be independent of the simulation volume and corresponds
to approximately $0.17\kMpc$. 
The ranking of the models is pretty much independent of $V$, with SPT always being the first to break down and EFT the last. 
However, RegPT does better than RLPT for small $V$ while the order is reversed for large $V$.
It is also worth noticing that, while RegPT quite significantly extends the reach of SPT for the power spectrum for $V\simeq 8 \cGpc$, it gives much smaller improvements for the bispectrum.

The nominal range of accuracy of EFT always extends beyond $\kfit$
(indicated with horizontal grey lines in the figures).
This is not surprising because, when the
$\chim$ statistic suggests a good fit at $\kfit$, our
definition of the reach will automatically pick a larger wavenumber.
Essentially, what this means is that
the EFT fits at $\kfit$ are good (or even too good) in terms of $\chim$.
We remind the reader that the values for $\kfit$ we use
are chosen 
in section~\ref{ssec:fitting_the_eft_params} based on two criteria:
(i) avoiding that the best-fitting EFT parameters run with $\kfit$ and (ii)
requiring consistency between the results obtained from $P$ and $B$.
However, since section~\ref{ssec:fitting_the_eft_params} takes into consideration the full \minerva set, our selected values might be considered `conservative' when $V$ is reduced (although we believe we should always perform the most challenging test for the theory, i.e. use the largest possible volume to test its basic assumptions like the scale-independence of the free parameters).
For comparison, 
in the top-left panels of Figs.~\ref{fig:p_k_vs_V_s1} 
and \ref{fig:fig_bisp_k_vs_V_s3},
we also show the range of accuracy one
would obtain by fitting the EFT parameters up to $\kfit=0.22\kMpc$
(yellow and green dashed lines).
This vastly increases the reach
at small $V$ (for both $P$ and $B$) but
reduces it at large $V$.
In particular, for large enough volumes, the estimated reach becomes smaller than $\kfit$ meaning that it is impossible to get a good fit to the numerical data. 

\begin{figure} 
	\centering
	\includegraphics[width=0.5\textwidth]{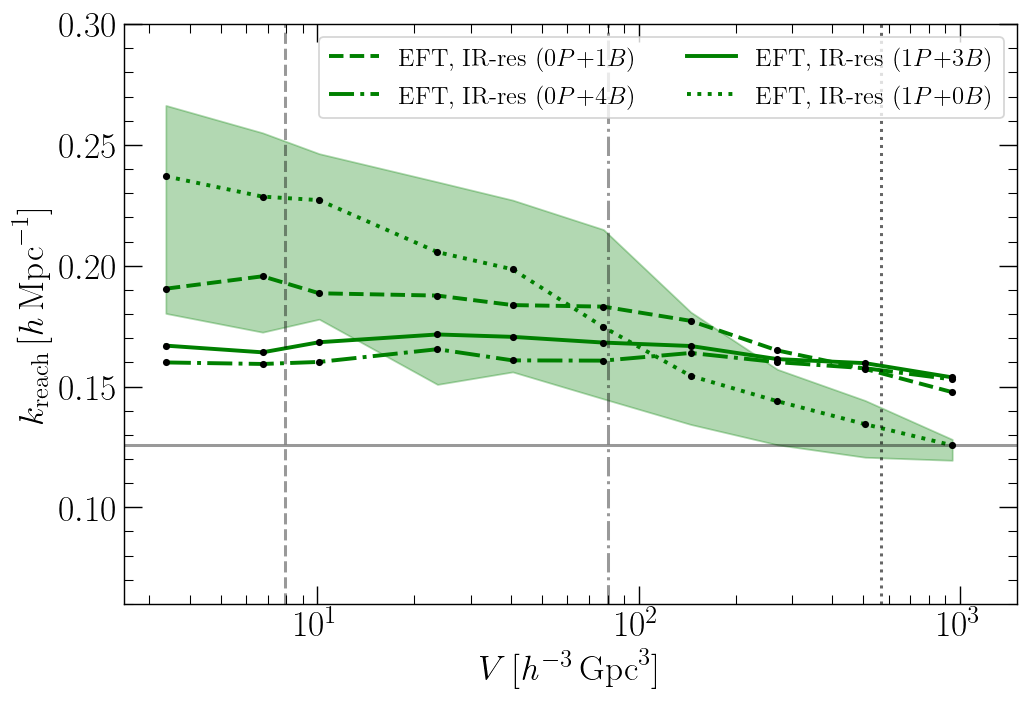}
	\caption{As in the top-left and top-right panels of Fig.~\ref{fig:fig_bisp_k_vs_V_s3} but for the \eftir{} model with the counterterms determined as described in  Table~\ref{Tab:counterfitmethods}.}
	\label{fig:kmax_vs_vol_all_params}
\end{figure}

So far we have concentrated on the median range of accuracy of each model. For this reason, in the top-right panels of Figs.~\ref{fig:p_k_vs_V_s1} and \ref{fig:fig_bisp_k_vs_V_s3},
we plot the statistical uncertainty of the estimated reach as a function of $V$. In this case,
we only consider SPT and IR-resummed EFT to improve readability.
The shaded areas indicate the central 68-per-cent region\footnote{Obviously, this statistic underestimates the actual scatter when $V$ approaches the total volume of the Minerva simulations as the different samples mostly overlap.} among the 200 sets of simulations with volume $V$. 
It turns out that the error on the reach is by no means negligible, particularly for EFT which contains free parameters. It is therefore important to take this into account when comparing studies based on different simulations.
In Fig.~\ref{fig:countervsV}, we show how the distribution of the best-fitting amplitudes for the counterterms varies with $V$. We consider the \eftir{} model for the bispectrum, fit $c_0$ from $P$ and the other counterterms from $B$, and use $\kfit=0.125\kMpc$.
It is important to notice that,
while the median values of the EFT parameters approximately coincide with those in Table~\ref{tab:counterfit}, the scatter
around them strongly depends on $V$. 
For the redshift shell in a Euclid-like survey, $c_1, c_2$ and $c_3$ show a tremendous variability meaning that they cannot be accurately measured from a single realisation.
In order to get a signal-to-noise ratio of order unity for them, it is necessary
to consider volumes
$V> 500 h^{-3}\mathrm{Gpc}^3$.
As a means to further investigate the impact of the
fitting strategy for the counterterms, 
in Fig.~\ref{fig:kmax_vs_vol_all_params}, we consider four methods
in which the EFT parameters are determined in different ways
as indicated in
Table~\ref{Tab:counterfitmethods}.
For the full \minerva data set, 
our standard choice ($1P+3B$) corresponds to the largest reach, while the one-parameter fit $1P+0B$
performs best for $V<50\cGpc$ suggesting
that there is no need to use three
counterterms when the surveyed volume
is small and the error bars of the measurements are large. 
 \citet{Angulo2015} reached similar conclusions using $V=27\cGpc$ and $\kfit=0.1\kMpc$
(conjecturing that the other counterterms give contributions comparable in size to two-loop corrections).
In Fig.~\ref{fig:kmax_vs_vol_all_params},
there is nothing surprising about the fact that models
with less free parameters can have a larger reach given that
the EFT counterterms are determined using $\kfit=0.125\kMpc$ and the estimated reach
is substantially larger than that.
It is interesting to try to understand why it is preferable
to set $c_1=c_2=c_3=0$ for small $V$. We believe that the reason is related to the fact that the expected values given in Table~\ref{tab:counterfit} are much smaller than the scatter seen in Fig.~\ref{fig:countervsV}. Basically, the fit picks large `random' counterterms in each realisation in order to adjust to the specific noise features. 

 In conclusion, the peculiarity of the EFT approach is the presence of free parameters in the counterterms that
need to be determined from the measurements. Our results show that
the methodology used to fix the EFT parameters heavily influences the range of accuracy of the theory. Basically, when the $V$ is small, error bars are large, and the counterterms are poorly determined, the resulting freedom in the
EFT parameters boosts the apparent
reach of the models.
Care should then be taken to ensure that
results from different studies are properly compared.
Moreover, future studies should carefully investigate if and how the freedom in the counterterms impacts the estimation of cosmological parameters from the galaxy bispectrum that will be measured by the forthcoming generation of redshift surveys \citep{Oddo21}.

\begin{table}
	\centering
	\bgroup
	\def\arraystretch{1}
	\begin{tabular}{c|c|c|c|c}
		\hline
		Method & $c_0$ & $c_1$ & $c_2$ & $c_3$\\
		\hline
	$1P+0B$ & $P$ & 0 & 0 & 0	\\
	$1P+3B$ & $P$ & $B$ & $B$ & $B$	\\
	$0P+1B$ & $B$ & 0 & 0 & 0	\\
	$0P+4B$ & $B$ & $B$ & $B$ & $B$	\\
		\hline
	\end{tabular}
	\egroup
	\caption{Schematic description of the methods used to fit the EFT counterterms in  Fig.~\ref{fig:kmax_vs_vol_all_params}. The symbols $P$ and $B$ denote parameters determined by fitting (up to $\kfit=0.125\kMpc$) the power spectrum or the the bispectrum, respectively. The number 0 indicates that the parameter is set to zero.
	}
	\label{Tab:counterfitmethods}
\end{table}

\subsection{Systematic errors}

Just like any other numerical method, 
$N$-body simulations do not provide the exact solution to the problem of gravitational instability and perturbation growth. Modern codes are optimised based on a
trade-off between computation speed and accuracy. Their finite mass and force resolution, the time-stepping criterion, the integration method, the way 
initial conditions are set and
forces are computed generate small systematic deviations from the exact solution.

Several studies try to quantify
the impact of these imperfections on various summary statistics
\citep[e.g.][]{TakahashiYoshida2008,Nishimichi2009,Baldauf2015,Schneider2016}.
However,
the current understanding is not mature enough yet
to provide a robust method for correcting goodness-of-fit statistics such as our $\chim$.
Therefore, simplified approaches are necessary. 
The most elementary consists of 
adding small uncorrelated systematic errors to the statistical error budget.
We follow this approach in the bottom-left panels of Figs.~\ref{fig:p_k_vs_V_s1} and
\ref{fig:fig_bisp_k_vs_V_s3}.
For the power spectrum, 
we add a $0.5$-per-cent systematic error in quadrature to the statistical error
in order to approximately match the numerical results of  \citet{Schneider2016}.
For the bispectrum, instead, we adopt two different approaches.
First, following \citet{Angulo2015}, we consider a shape- and scale-independent systematic contribution at the 2-per-cent level (again summed in quadrature to the random error).
As a second option, we use
the scale-dependent ansatz 
by \citet{Baldauf2015} which provides a fit to the systematic deviations measured among $N$-body simulations
with different characteristics.
In this case, the (relative) systematic error is
\begin{equation}
\label{eq:baldauf_error}
    \frac{\Delta B}{B} = 0.01 + 0.02 \left(\frac{k_1}{0.5\kMpc}\right)\, ,
\end{equation}
where, as always, $k_1$ denotes the largest side of the triangular configuration. 
Since both approaches give very similar results, in Fig.~\ref{fig:fig_bisp_k_vs_V_s3} we only show those obtained
with the scale-dependent ansatz.
We are now ready to present our findings.
For the models with no free-parameters,
adding small systematic errors only changes the reach for large values of $V$, i.e. in every case in which the statistical errors
are smaller than the additional systematic contributions.
As a consequence,
the resulting ranges of accuracy show
little variations with $V$ and
RegPT turns out to consistently have the largest reach for all values of $V$.
Conversely, the range of accuracy of EFT
is strongly affected by the inclusion of systematic errors for all $V$.
What is perhaps more surprising is that the reach of the EFT models increases with $V$. This happens because
the values assigned to the EFT parameters scatter among the 200
subsets of simulations. In particular, when $V$ is small, the EFT parameters have big uncertainties and the models cannot provide a good fit to the numerical data for large $\kmax$.
To clarify this further, we investigate
what happens when we use the same EFT
parameters for all simulation subsets. In this case, we use the mean of the values obtained from the individual sets. Our results are shown with dot-dashed lines in the bottom-left panels of Figs.~\ref{fig:p_k_vs_V_s1} and
\ref{fig:fig_bisp_k_vs_V_s3}. 
The reach for the power spectrum and bispectrum are extended to roughly $0.43$ and $0.26\kMpc$, respectively, 
independently of $V$.
Note, however, that the calibration of the EFT parameters without using the data is not doable in practical applications to observational surveys.
%%%%%%% Resolution error

Systematic errors affect the accuracy and not the precision of measurements.
Therefore, it is somewhat unnatural to model them as random uncorrelated errors.
For example, it is well known that the finite mass resolution
of $N$-body simulations leads to the suppression of density fluctuations on small scales \citep[ e.g.][]{Heitmann+2010,Schneider2016}. 
In what follows, we propose a simple parameterisation of this effect which
allows us to include it in our error budget as a `perfectly correlated' error.
Let us consider an $N$-body simulation
with particle density $\bar{n}$ and
make the educated guess that, due to the finite mass resolution,
$\delta_{\bar{n}}(\kv)=\delta_{\infty}(\kv)\,R_{\bar{n}}(k)$ where $\delta_{\infty}(\kv)$ denotes the
ideal continuum case. It follows that
\be
P_{\bar{n}}(k)=P_{\infty}(k)\,R^2_{\bar{n}}(k)\,,
\ee
and 
\be
\label{eq:Bn}
B_{\bar{n}}(k_1,k_2,k_3)=B_{\infty}(k_1,k_2,k_3)\,R_{\bar{n}}(k_1)\,R_{\bar{n}}(k_2)\,R_{\bar{n}}(k_3)\,.
\ee
In order to constrain the shape of
the function $R_{\bar{n}}(k)$ at $z=1$,
we use the power spectra extracted 
from 30 realisations of the Quijote simulations \citep{QuijoteSims2019} at three different resolutions, i.e. using $256^3$, $512^3$ and $1024^3$ particles
within a box of 1 $h^{-3}\, \mathrm{Gpc}^3$.
We find that the following parameterisation:
\be
\label{eq:r}
R_{\bar{n}}(k)=\frac{1}{1+A(\bar{n})\,\epsilon(k)}
\ee
with 
\be
\epsilon(k)=\frac{k}{\kMpc}+\alpha\,\left(\frac{k}{\kMpc}\right)^2+\beta\,\left(\frac{k}{\kMpc}\right)^3\,,
\ee 
accurately reproduces the numerical data
up to $\kmax=0.3\kMpc$.
The ratio between two power spectra obtained
with different mass resolution is
\be
\frac{P_{\bar{n}_1}(k)}
{P_{\bar{n}_2}(k)}=\frac{R^2_{\bar{n}_1}(k)}{R^2_{\bar{n}_2}(k)}
\simeq 1+2\,\Delta A\,\epsilon(k)\;,
\ee
where $\Delta A=A(\bar{n}_2)-A(\bar{n}_1)$ and we have Taylor expanded the final result to first order assuming that the corrections are small on the scales of interest. From the Quijote power spectra, we obtain
$\alpha=-0.35$ and $\beta=0.39$.
In order to estimate $A(\bar{n}_\minerva)$, we assume that
the correction is negligible at the
highest Quijote resolution and
interpolate $\Delta A$ (note that $\bar{n}_\minerva=0.296 \,h^{3}\,\mathrm{Mpc}^{-3}$ while
the Quijote simulations have $\bar{n}=0.017,0.134,$ and $1.074 \,h^{3}\,\mathrm{Mpc}^{-3}$). 
We obtain $A(\bar{n}_\minerva)=0.0188$
which corresponds to sub-per-cent corrections over all scales of interest.
Eventually, we write the systematic
error (bias) in the summary statistics extracted from an $N$-body simulation as
\begin{align}
\label{eq:delta_P_res}
\Delta P  &=   P_{\bar{n}}- P_\infty=
P_{\bar{n}}\,\left(1-\frac{1}{R^2_{\bar{n}}(k)} \right)
\approx -2\,P_{\bar{n}}\,A(\bar{n})\,\epsilon(k)\,,\\
\label{eq:delta_B_res}
\Delta B  &=   B_{\bar{n}}-B_\infty\approx -B_{\bar{n}}\, A(\bar{n})\,\left[\epsilon(k_1)+\epsilon(k_2)+\epsilon(k_3)\right]\,.
\end{align}
In order to apply this result to the
\minerva simulations we re-scale
our estimate for $A(\bar{n}_\minerva)$
by a factor $\gamma$ for which
we consider three possible values, namely 0.5, 1.0 and 1.5.
Although these
values may not exactly describe the correction due to the finite mass resolution in the \minerva runs, they
allow us to
conceptually investigate the effect of a scale- and shape-dependent bias.
We thus re-compute the $\chim$ statistic after shifting the measurements from the simulations according to 
the corrections given in equations (\ref{eq:delta_P_res}) and (\ref{eq:delta_B_res}). Our results 
for the reach of the models
are displayed in the bottom-right panels of Figs.~\ref{fig:p_k_vs_V_s1} and \ref{fig:fig_bisp_k_vs_V_s3} where the coloured bands indicate the range of 
variability induced by $\gamma$ and
the solid lines reproduce the curves
from the top-left panel to emphasize changes.
Overall, the impact of the corrections is rather minor. Nonetheless, a few changes are worth noticing.
For the power spectrum, the \eftir{} at intermediate volumes shows
the most marked improvement.
For the bispectrum, accounting for the bias improves the reach of RegPT for large $V$ and deteriorates it for both EFT models. Perhaps, the most important conclusion that one can draw from this test is that its results are very different from those obtained by simply inflating the random errors to account for systematics (as routinely done in the literature). The latter approach, in fact, artificially boosts the reach of models with free parameters as in the case of EFT. Our study calls for a better understanding of random and systematic errors in $N$-body simulations.

\section{Summary}
\label{sec:conclusion}

Perturbative techniques based on fluid dynamics are widely used to study the growth of the large-scale structure of the Universe.
In fact,
they often are the only method of
obtaining predictions with analytical control.
The convergence properties of perturbation theory are still a matter of debate but there is mounting evidence that the resulting expressions for
large-scale observables 
are actually asymptotic,
i.e. only the truncated series
expansion (including just the first few terms)
provides an accurate approximation to the exact solution
\citep[e.g.][and references therein]{Pajer-vanderWoude-2018, Konstandin+2020}.

Modern perturbative approaches 
come in a plethora of flavours and
sometimes contain free parameters. 
It is thus imperative to identify their regime of validity and accuracy before applying them to practical situations.
$N$-body simulations of collisionless dark matter in a cosmological background are the standard test bed for inferring the reach of the different models.

The purpose of this study is threefold.
First, we use a very large set of $N$-body simulations (the \minerva suite) to test the
NLO expansions for the matter power spectrum and bispectrum in five different implementations of perturbation theory, namely SPT, RegPT, RLPT, EFT and \eftir{}.
Second, we try to draw the line that demarcates general results from those affected by the method used to determine the
reach of the models with free
parameters (i.e. EFT and \eftir{}).
Third, we explore a novel way to account for the systematic errors introduced by the finite mass resolution of $N$-body simulations.

Specifically, we study how well the different models match the measurements from the simulations as a function of the maximum wavenumber considered, $\kmax$.
Having in mind the forthcoming
generation of surveys such as those that will be conducted by DESI and the Euclid mission, we only consider data at $z=1$.
We define the reach of a model as the minimum $\kmax$ at which the $\chi^2$ goodness-of-fit test rejects the null hypothesis that the $N$-body data are consistent with the model predictions at the significance level of 0.05.
This requires making some assumptions about the covariance matrix of the measurements. We use the Gaussian
approximation given in equation~(\ref{eq:Cpp})
for the power spectrum and
a more sophisticated expression for the bispectrum -- see equation~(\ref{eq:C_BB}). In both cases, we use a dedicated version of the $\chi^2$ test to verify that these expressions closely approximate the covariance matrix of
the measurements extracted from the simulations. 

In the first part of our study, we
consider the full \minerva suite
and neglect systematic errors in the simulations.
Our main findings are as follows.
\begin{enumerate}[(i),wide, nosep]
\item 
By fitting the EFT parameters that determine the amplitude of the counterterms to the simulation data
as a function of $\kmax$, we find 
that they remain stable until 
a maximum wavenumber and change beyond that
(Figs. \ref{fig:eft_ps_fit} and \ref{fig:eft_bisp_fit}). 
The stability region ends at $\kmax=0.14\kMpc$ for the power spectrum
and $0.125\kMpc$ for the bispectrum.
We use these values to define the default range of scales ($k<\kfit$) over which we fit the EFT parameters.
\item
The $\chi^2$ goodness-of-fit test for the power spectrum (Fig.~\ref{fig:ps_chi_s1})
shows that EFT and \eftir{} accurately
match the simulations up to 
$\kmax=0.14 \kMpc$
while all the models without free parameters fail at much larger scales, i.e. $\kmax=0.06\kMpc$.
\item
Repeating the test for the bispectrum (Fig.~\ref{fig:bs_chi_s1}) 
provides a clear ranking for the models based on their reach.
The EFT models have the largest range of accuracy ($\kmax\simeq 0.16$ - $0.19\kMpc$, depending on the binning of the data) followed by RegPT and RLPT
($\kmax\simeq 0.10$ - $0.14\kMpc$)
and SPT ($\kmax\simeq 0.08\kMpc$).
Note that the nominal reach of EFT extends beyond $\kfit$, meaning that the model with the
counterterms fixed using triangle configurations with $k<\kfit=0.125\kMpc$
continues to provide a good fit on (slightly) smaller scales.
\end{enumerate}

Next, by sub-sampling the \minerva suite, we investigate how the reach of the models depends on the total volume covered by the simulations used in our tests. This is particularly useful when comparing different results in the literature and
also to gauge the range of scales that can be robustly probed in an actual galaxy redshift survey.
In this analysis, we approximately account for systematic effects introduced by the $N$-body technique using different methods.
Our key results are as follows.
\begin{enumerate}[(i),wide, nosep]
 \setcounter{enumi}{3}
 \item Obviously, the reach of the models improves for smaller volumes
 as the statistical error bars become larger and it is easier to fit the data. Considering a 
 redshift bin of width $\Delta z=0.2$
 centred at $z=1$ for a Euclid-like survey, gives a median reach for SPT of approximately $0.12 \kMpc$ for the power spectrum and $0.15 \kMpc$ for the bispectrum. On the other hand, for \eftir{} we obtain 
 $0.25 \kMpc$ for the power spectrum and $0.18 \kMpc$ for the bispectrum. All the other models lie in between these extremes (Figs.~\ref{fig:p_k_vs_V_s1} and
\ref{fig:fig_bisp_k_vs_V_s3}). 
It is also important to mention that
the scatter of the reach between different realisations with the same volume becomes rather large for the
models that have free parameters (the central 68-per-cent range for EFT extends from $0.19$ to $0.34\kMpc$ in the case of the power spectrum). This should be taken into account when comparing results from different studies.
 \item The estimated range of accuracy of the EFT predictions is heavily influenced by the procedure adopted to fit the counterterms. For the volume of the Euclid-like shell,
 using $\kfit=0.22 \kMpc$ extends the median reach of the \eftir{} model to $0.33$
 and $0.25\kMpc$ for the power spectrum and the bispectrum, respectively,
 but degrades it for the full \minerva set.
 For the bispectrum, fitting only $c_0$ from the power spectrum and setting the other three counterterms to zero gives the largest reach for $V<100\,h^{-3} \mathrm{Mpc}^3$. Fitting all the four parameters is instead preferred for larger volumes  (Fig.~\ref{fig:kmax_vs_vol_all_params}). Therefore, it is difficult to unequivocally define a reach for the models with free parameters.
\item The results above are only slightly affected (less than 10 per cent change) by accounting for a scale-
and shape-dependent bias due to the
finite mass resolution of the $N$-body simulations. 
\item The situation is very different when uncorrelated
systematic errors are added in quadrature to the statistical uncertainties, as assumed in \citet{Baldauf2015} and \citet{Angulo2015}. In this case, the reach of EFT is dramatically extended thanks to the freedom provided by the
counterterms. For example, considering the whole \minerva suite, 
we obtain that the \eftir{} model
provides a good fit until
0.40 and $0.27\kMpc$ for the power-spectrum and the bispectrum, respectively.
More modest changes are seen for the models with no fixed parameters at large $V$.
\end{enumerate}

In order to constrain the cosmological parameters from the galaxy bispectrum, it is necessary to model galaxy biasing, discreteness effects, and redshift-space distortions on top of the non-linearities of the matter density field.
It is very well possible that
the additional terms in the expressions for the galaxy bispectrum to NLO
will be degenerate with higher-order
terms in the matter models
and thus extend the reach
of the more complex mathematical descriptions beyond the scales determined in this work.
Yet, it is pivotal to retain control over the extent to which this is happening, especially if one wants to assign a physical meaning to the additional (e.g. bias and shot-noise) parameters.
This is why we believe our work is important.

\section*{Acknowledgements}

We are grateful to Claudio Dalla Vecchia and Ariel Sanchez for running and making available the \minerva simulations, performed and analysed on the Hydra and Euclid clusters at the Max Planck Computing and Data Facility (MPCDF) in Garching. 
We thank Martin Crocce, Alex Eggemeier, Azadeh Moradinezhad, Roman Scoccimarro and Zvonimir Vlah  for useful discussions.
We acknowledge the hospitality of the Institute for Fundamental Physics of the Universe in Trieste
where part of this work was carried out in October 2019.
D.A. acknowledges partial financial support by the Shota Rustaveli National Science Foundation of Georgia (GNSF) under the grant FR-19-498.
E.S. is partially supported by the INFN INDARK PD51 grant and acknowledges support from PRIN MIUR 2015 Cosmology and Fundamental Physics: illuminating the Dark Universe with Euclid.
M.B. acknowledges support from the Netherlands Organization for Scientific Research (NWO), which is funded by the Dutch Ministry of Education, Culture and  Science  (OCW),  under  VENI  grant  016.Veni.192.210. M.B. also acknowledges support from the NWO under the project ``Cosmic Origins from Simulated Universes'' for the computing time allocated to run a subset of the \textsc{Eos} Simulations on \textsc{Cartesius}, a supercomputer which is part of the Dutch National Computing Facilities.
A.L. acknowledges funding by the LabEx ENS-ICFP: ANR-10-LABX-0010/ANR-10-IDEX-0001-02 PSL*.
V.Y. acknowledges funding from the European Research Council (ERC) under the European Union's Horizon 2020 research and innovation programme (grant agreement No 769130).

\section*{Data availability}
The data underlying this article will be shared on reasonable request to the corresponding author.
%%%%%%%%%%%%%%%%%%%%%%%%%%%%%%%%%%%%%%%%%%%%%%%%%%



\bibliographystyle{mnras}
\bibliography{refs.bib} % if your bibtex file is called example.bib

%%%%%%%%%%%%%%%%%%%%%%%%%%%%%%%%%%%%%%%%%%%%%%%%%%

%%%%%%%%%%%%%%%%% APPENDICES %%%%%%%%%%%%%%%%%%%%%

\appendix
\section{SPT}
\label{appendix:spt}
The one-loop correction to the matter power spectrum in SPT is
\begin{align}
   P^\text{1-loop}_\text{SPT}(k,z)=P_{13}(k,z)+P_{22}(k,z) \, ,
\end{align}
where
\begin{align}
    &P_{13}(k,z)=6 \, [D(z)]^4 \, \Pl(k) \int_\qv F_2(\kv,\qv,-\qv)\,\Pl(q) \, ,\\
    &P_{22}(k,z)=2 \, [D(z)]^4 \, \int_\qv [F_2(\qv,\kv-\qv)]^2 \,\Pl(|\kv-\qv|)\,\Pl(q) \, ,
\end{align}
and $\int_{\textbf{q}}$ denotes $\int \frac{\dd^3 \mathbf{q}}{(2\pi)^3}$.

Similarly, for the bispectrum, we have:
\begin{align}
    B^\text{1-loop}_\text{SPT} = B_{222}+B_{321}&^{I}+B_{321}^{\,\,\,\,\,\,\,\,II}+B_{411}\, ,
\end{align}
with
\begin{align}
\label{eq:b222}
B_{222}&\left(k_1,k_2,k_3, z\right)=8\,[D(z)]^6\,\int_{\textbf{q}}\Pl\left(q\right)\Pl\left(|\textbf{k}_2-\textbf{q}|\right)\Pl\left(|\textbf{k}_3+\textbf{q}|\right)  \nonumber \\
&\times F_2\left(-\textbf{q},\textbf{k}_3+\textbf{q}\right) 
%\nonumber \\
F_2\left(\textbf{k}_3+\textbf{q},\textbf{k}_2-\textbf{q}\right)F_2\left(\textbf{k}_2-\textbf{q},\textbf{q}\right)
\end{align}
\begin{align}
\label{eq:b321i}
B_{321}&^{I}\left(k_1,k_2,k_3, z\right) = 6\,[D(z)]^6\,\Pl\left(k_3\right)\int_{\textbf{q}}\Pl\left(|\textbf{k}_2-\textbf{q}|\right) \Pl\left(q\right) \nonumber \\
&\times F_3\left(-\textbf{q},-\textbf{k}_2+\textbf{q},-\textbf{k}_3\right)F_2\left(\textbf{k}_2-\textbf{q},\textbf{q}\right)+ \text{5 perms.}
\end{align}
\begin{align}
\label{eq:b321ii}
B_{321}&^{II}(k_1,k_2,k_3, z) = 6\, [D(z)]^6\, \Pl(k_2) \Pl(k_3)F_2^{(s)}(\textbf{k}_2,\textbf{k}_3)  \nonumber \\
&\times \int_{\textbf{q}}\Pl\left(q\right)F_3\left(\textbf{k}_3,\textbf{q},-\textbf{q}\right) + \text{5 perms.}
\end{align}
\begin{align}
\label{eq:b411}
B_{411}&\left(k_1,k_2,k_3, z\right)=12\,[D(z)]^6\,\left(z\right)\Pl\left(k_2\right)\Pl\left(k_3\right)  \nonumber \\
&\times \int_{\textbf{q}}\Pl\left(q\right)F_4\left(\textbf{q},-\textbf{q},-\textbf{k}_2,-\textbf{k}_3\right) + \text{2 perms.} \, .
\end{align}

%%%%%%%%%%%%%%%%%%%%%%%%%%%%%%%%%%%%%
\section{RegPT}
\label{appendix:rpt}
The $(p+1)$-point propagator, $\Gamma^{(p)}(\mathbf{k}_1,\dots,\mathbf{k}_p,z)$, is defined as
\begin{align}
\label{eq:gamma_def}
\frac{1}{p!}\Bigg\langle \frac{\delta^p \delta(\mathbf{k},z)}{\delta\deltainit(\kv_1)\cdots\delta\deltainit(\mathbf{k}_p)}\Bigg\rangle &= \frac{\delta_\mathrm{D}(\mathbf{k}-\mathbf{k}_{1\cdots p})}{(2 \pi)^{3(p-1)}}\,\Gamma^{(p)}\,,
\end{align}
and can be expanded
using equations (\ref{eq:psi_n}) and (\ref{eq:gamma_def}) as
\begin{align}
\label{eq:gamma_spt_expansion}
\Gamma^{(p)}=\Gamma^{(p)}_{\mathrm{tree}}+\sum_{n=1}^{\infty}\Gamma^{(p)}_{n-\text{loop}} \, .
\end{align}
$\Gamma^{(p)}_{\text{tree}}(\mathbf{k}_1,\dots,\mathbf{k}_p,z)=[D(z)]^p\, F_p(\mathbf{k}_1,\dots,\mathbf{k}_p)$ and 
\begin{align}
\nonumber
\label{eq:gamma_n_loop}
\Gamma&^{(p)}_{n-\text{loop}}(\mathbf{k}_1,\dots,\mathbf{k}_p,z)=[D(z)]^{(2n+p)}\,C^{2n+p}_p\,
(2n-1)!! \\
\nonumber
& \times\int \frac{\dd^3\mathbf{q}_1 \cdots \dd^3\mathbf{q}_n}{(2\pi)^{3n}} F_{2n+p}(\mathbf{q}_1,-\mathbf{q}_1,\dots,\mathbf{q}_n,-\mathbf{q}_n,\mathbf{k}_1,\dots,\mathbf{k}_p) \\
&\times \Pl(q_1)\cdots \Pl(q_n)
\equiv [D(z)]^{(2n+p)} \,\overline{\Gamma}^{\,(p)}_{n-\text{loop}}(\mathbf{k}_1,\dots,\mathbf{k}_p)\,,
\end{align}
where $C^{2n+p}_p$ denotes the binomial coefficient. 
Resumming the subset
of terms that provide the dominant
contribution at small scales gives \citep{2006PhRvD..73f3520C, Bernardeau2008}
\begin{equation}
\label{eq:gamma_high_k}
\Gamma^{(p)}(\mathbf{k}_1, \dots,\mathbf{k}_p,z)\xrightarrow{k\rightarrow\infty}\exp\left\{-\frac{k^2\, [D(z)]^2\,\sigma^2_\mathrm{d}}{2}\right\}\,\Gamma^{(p)}_{\text{tree}} \, .
\end{equation}
where 
\begin{equation}
\sigma_\mathrm{d}^2=\frac{1}{3}\int \frac{\dd^3 \mathbf{q}}{(2\pi)^3}\frac{\Pl(q)}{q^2} 
\end{equation}
is the rms value of the one-dimensional linear displacement field.
Up to one-loop order, the regularized propagators  
which interpolate between the two asymptotic regimes are \citep{Taruya2012, PhysRevD.85.123519}
 \begin{align}
 \nonumber
\Gamma^{(1)}_{\mathrm{reg}}(k,z)=D(z)\ &\left\{1+\frac{k^2\,[D(z)]^2\, \sigma_\mathrm{d}^2 }{2} +[D(z)]^2 \,\overline{\Gamma}^{\,(1)}_{1-\text{loop}}(k)\right\} \\
&\times\exp\left\{-\frac{k^2\, [D(z)]^2\,\sigma^2_\mathrm{d}}{2}\right\}
\end{align}
 \begin{align}
 \nonumber
\Gamma^{(2)}&_{\mathrm{reg}}(\mathbf{k}_1,\mathbf{k}_2,z)=[D(z)]^2\, \left(F_2(\mathbf{k}_1,\mathbf{k}_2)\,\left\{1+\frac{k^2\,[D(z)]^2\, \sigma_\mathrm{d}^2}{2}\right\} \right. \\ 
&+ [D(z)]^2 \,\overline{\Gamma}^{\,(2)}_{1-\text{loop}}(\kv_1,\kv_2)\Bigg)\,
\exp\left\{-\frac{k^2\, [D(z)]^2\,\sigma^2_\mathrm{d}}{2}\right\}\, ,
\end{align}
 \begin{align}
 \nonumber
\Gamma^{(3)}_{\mathrm{reg}}(\mathbf{k}_1,\mathbf{k}_2,\mathbf{k}_3,z)=[D(z)]^3\, F_3&(\mathbf{k}_1,\mathbf{k}_2,\mathbf{k}_3) \\
&\times \exp\left\{-\frac{k^2\, [D(z)]^2\,\sigma^2_\mathrm{d}}{2}\right\}\,.
  \end{align}

In this formalism,
the matter power spectrum and bispectrum up to one-loop corrections can be expressed as~\citep{Bernardeau2008}
\begin{align}
\label{eq:power_spectrum_one_loop_gamma}
\nonumber
P(k,z) =  [\Gamma^{(1)}&(k,z)]^2 \ \Pl(k)\\&+ 2 \int_{\textbf{q}} [\Gamma^{(2)}(\qv, \kv-\qv,z)]^2 
\,\Pl(q)\, \Pl(|\kv-\qv|) \, ,
\end{align} 
\begin{align}
\label{eq:bispectrum_one_loop_gamma}
\nonumber
B(\kv_1,&\kv_2,\kv_3,z) = 2 \, \Gamma^{(2)}(\mathbf{k}_1,\mathbf{k}_2,z) \ \Gamma^{(1)}(k_1,z) \ \Gamma^{(1)}(k_2,z) \nonumber \\
&\times \Pl(k_1) \ \Pl(k_2) +  \text{2 perms.} \nonumber \\&+\bigg[8 \int_{\textbf{q}} \Gamma^{(2)}(\mathbf{k}_1+\mathbf{q},-\mathbf{q},z)\,  \Gamma^{(2)}(-\mathbf{k}_1-\mathbf{q},\mathbf{q}-\mathbf{k}_2,z)\nonumber\\
& \times \Gamma^{(2)}(\mathbf{k}_2-\mathbf{q},\mathbf{q},z) \Pl(|\mathbf{k}_2-\mathbf{q}|)
\Pl(|\mathbf{k}_1+\mathbf{q}|) \Pl(q)\bigg] \nonumber\\&+ \bigg[6 \ \Gamma^{(1)}(\mathbf{k}_3,z)\,\Pl(k_3) 
\int_{\textbf{q}} \Gamma^{(3)}(\mathbf{q}-\mathbf{k}_2,-\mathbf{k}_3,-\mathbf{q},z) \nonumber\\ &\times \Gamma^{(2)}(\mathbf{q},\mathbf{k}_2-\mathbf{q},z) 
\Pl(|\mathbf{k}_2-\mathbf{q}|)\Pl(q)+ 5 \,\text{perms.}\bigg]\, .
\end{align}

 \section{RLPT}
 \label{appendix:rlpt}

By combining the Lagrangian perturbative expansion

with equations~\eqref{eq:RLPT_P1} and \eqref{eq:RLPT_B1}, one obtains the following expressions at one loop
\citep{Matsubara2008,Rampf2012b}:
\begin{align} \label{eq:RLPT_B2}
\nonumber
P_{\mathrm{RLPT}}(k) &=  \left[  \Pl + P_{\mathrm{SPT}}^{\text{1-loop}} + \frac{k^2}{6 \pi^2} \Pl  \int \dd q \, \Pl (q)  \right]  \\
& \times \exp \left[ -\frac{k^2}{6 \pi^2} \int \dd q \, \Pl (q) \right] \, , \\
\nonumber
B_{\mathrm{RLPT}}(&k_1, k_2, k_3) = \Bigg[  B_{\mathrm{SPT}}^{\mathrm{tree}} \Bigg\{ 1 + \frac{k_1^2 + k_2^2 + k_3^2}{12 \pi^2} \int \dd q \, \Pl (q)\Bigg\}   \\
 & +B_{\mathrm{SPT}}^{\text{1-loop}} \Bigg] \exp \left[ -\frac{k_1^2 + k_2^2 + k_3^2}{12 \pi^2} \int \dd q \, \Pl (q) \right] \, .
\end{align} 
\end{document}